\documentclass[aps,twocolumn,pra,10pt, showpacs,superscriptaddress,groupedaddress]{revtex4-2} 

\usepackage{subcaption}
\usepackage{braket}
\usepackage[abs]{overpic}

\usepackage{physics}
\usepackage{graphicx}
\usepackage{setspace}
\usepackage{bbold}
\usepackage{comment}
\usepackage{color}
\usepackage{soul}
\usepackage[abs]{overpic}
\usepackage{amsmath}
\usepackage{lipsum}
\usepackage{placeins}
\usepackage[abs]{overpic}
\usepackage{float}
\usepackage[export]{adjustbox}
\usepackage{amsmath}
\usepackage{nicefrac}
\usepackage{cleveref}
\usepackage{layouts}
\usepackage[font=small, labelsep=period,
   justification=Justified,
   format=plain]{caption} 
\usepackage{ragged2e}

\begin{document}
\setlength{\textfloatsep}{5pt}


\title{Synchronization in coupled laser arrays with correlated and uncorrelated disorder}

\author{Amit Pando}
\affiliation{ Department of Physics of Complex Systems, Weizmann Institute of Science, Rehovot 7610001, Israel}
\author{Sagie Gadasi}
\affiliation{ Department of Physics of Complex Systems, Weizmann Institute of Science, Rehovot 7610001, Israel}
\author{Eran Bernstein}
\affiliation{ Department of Physics of Complex Systems, Weizmann Institute of Science, Rehovot 7610001, Israel}
\author{Nikita Stroev}
\affiliation{ Department of Physics of Complex Systems, Weizmann Institute of Science, Rehovot 7610001, Israel}
\author{Asher Friesem}
\affiliation{ Department of Physics of Complex Systems, Weizmann Institute of Science, Rehovot 7610001, Israel}
\author{Nir Davidson}
\affiliation{ Department of Physics of Complex Systems, Weizmann Institute of Science, Rehovot 7610001, Israel}


\begin{abstract}
The effect of quenched disorder in a many-body system is experimentally investigated in a controlled fashion. It is done by measuring the phase synchronization (i.e. mutual coherence) of 400 coupled lasers as a function of tunable disorder and coupling strengths. The results reveal that correlated disorder has a non-trivial effect on the decrease of phase synchronization, which depends on the ratio of the disorder correlation length over the average number of synchronized lasers. The experimental results are supported by numerical simulations and analytic derivations.

\end{abstract}

\maketitle

\textit{Introduction - } 
Many different quantum and classical physical systems can be described by the framework of many-body interacting oscillators. Examples include transverse-field spin models, wherein spins rotating around a local magnetic field can synchronize to reach finite magnetization even in the presence of a spatially varying magnetic field \cite{kiely2018relationship,henkel1984statistical,juhasz2014random,mckenzie1996exact,fisher1992random}. Synchronization of classical phase oscillators has been studied for decades through the Kuramoto model \cite{acebron2005kuramoto}, and is manifested in many different systems such as arrays of Josephson junctions \cite{wiesenfeld1998frequency,trees2005synchronization}, coupled laser arrays \cite{yeung1999time,takemura2021emulating}, and even human networks \cite{shahal2020synchronization}. In all of these, disorder plays a major role in synchronization. There are cases where disorder leads to synchronization \cite{niederberger2010disorder,villain1980order}, but in general it acts as an obstacle, preventing the interaction between the individual members of the ensemble so they cannot synchronize.

While many theoretical investigations of the effects of disorder on synchronization have been performed  \cite{acebron2005kuramoto,sonnenschein2013approximate, granato1986quenched,vlasov2013synchronization,rouzaire2021defect}, it is difficult to verify the results experimentally because excessively high control and accuracy are needed. Several systems with robustness to disorder were recently reported \cite{contractor2022scalable,alex2021,bandres2018topological,rosiek2023observation,wang2009observation,hafezi2011robust}, but with limited ability to tune the disorder and quantify their robustness. 

In this work, we resort to an array of 400 lasers with nearest-neighbor coupling and well-controlled quenched (time-independent) disorder to obtain a tunable system for investigating the effects of disorder on synchronization (locking) of their optical phase. The disorder is introduced in the form of frequency detuning, where the resonant frequency of each individual laser is shifted. By precisely controlling the magnitude of the disorder, we show how it gradually diminishes the ability of the lasers to synchronize. By varying spatial properties of the disorder, we demonstrate how its effects depend on a non-trivial interplay between the scales of the problem, namely the correlation length of the disorder and the average number of synchronized lasers, in good agreement with our numerical and theoretical derivations.  

Our experimental system of coupled laser arrays can be readily extended to investigate the effects of controlled disorder on topological states \cite{vishwatopo,yang2022topological,alex2021}, non-Hermitian dynamics \cite{ruter2010observation,arwas2022anyonic}, geometric frustration \cite{nixon2013}, spin simulators and physical solvers for complex problems \cite{wang2013coherent,mcmahon2016fully,marandi2014network,tradonsky2019rapid}. 

\begin{figure}[h]
    \centering
    \setlength\fboxrule{0.2pt}
    \includegraphics[trim=4cm  3cm 1.5cm 9.5cm,clip,width=1.1\linewidth]{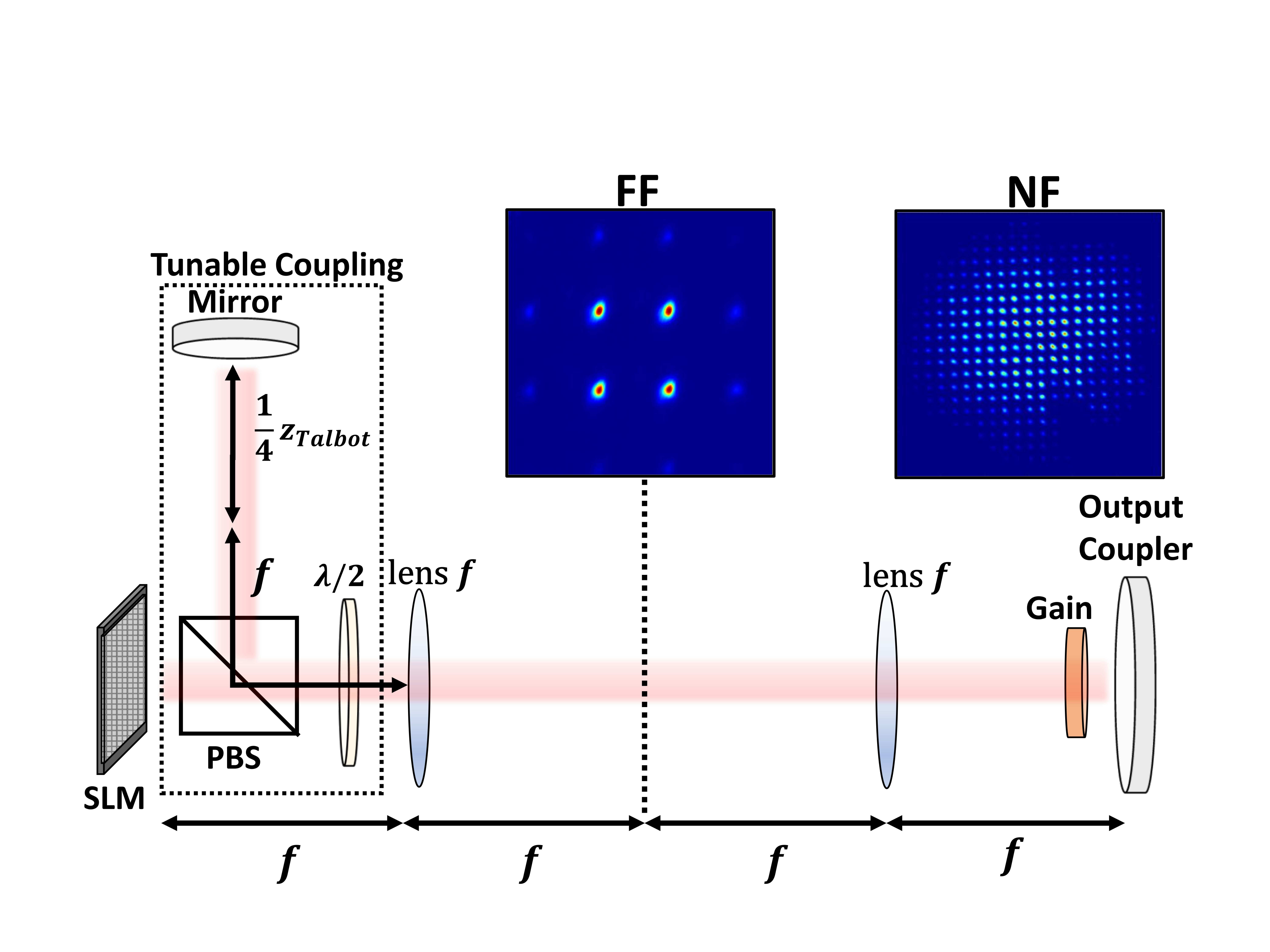}
    \caption{Experimental system. A modified digital degenerate cavity laser \cite{Tradonsky:21, cao2019complex, Arnaud:69} with an intra-cavity SLM that defines 400 lasers (20 by 20 square lattice) with a normally distributed random frequency detuning pattern having standard deviation $\tau_c\Omega_{RMS}$ and correlation length $\xi$.  The tunable coupling arrangement (surrounded by the dashed lines) introduces Talbot coupling \cite{tradonsky2017talbot,talbotLeger} between nearest neighbor lasers whose strength can be continuously tuned by rotating the $\lambda/2$ waveplate. The insets show representative far-field (FF, left) and near-field (NF, right) intensity distributions for zero disorder strength ($\tau_c\Omega_{RMS} = 0$). The four sharp diffraction FF peaks indicate near-perfect phase synchronization with $\pi$ phase difference between neighbors due to negative coupling.}
    \label{fig_exp_setup}
\end{figure}

\textit{Experimental System - } Our experimental system, schematically shown in Fig. \ref{fig_exp_setup} and described in detail in \cite{Pando:23} is comprised of a digital degenerate cavity laser (DDCL)\cite{Tradonsky:21,cao2019complex,Arnaud:69}. It includes an intra-cavity 4f telescope, a 98\% reflectivity output coupler, a 3mm thick ND:YVO4 gain medium lasing at wavelength $\lambda=1.06 \mu m$, a reflective spatial light modulator (SLM) with pixel size of $8\mu m$, and a tunable coupling arrangement. The gain medium has a fluorescence lifetime of $\tau_f\approx 100\mu s$, and is end-pumped by a $808nm$ diode laser with a pulse duration of $500\mu s$ at 4Hz repetition rate. The intra-cavity SLM forms a digital amplitude and phase mask, to form 400 independent lasers in a 20 by 20 square array with spacing between adjacent lasers of $d_{lat}=300\mu m$, (see NF inset in Fig. \ref{fig_exp_setup}) and to precisely control the frequency detuning between the lasers. 
By changing the phase retardation of each SLM pixel, we locally vary the effective cavity length with a precision of $\frac{\lambda}{256}$ and thereby detune the resonant frequency of each laser with a precision of $\tau_c\Delta\Omega=\frac{2\pi}{256} \text{rad}$, where $\tau_c=\frac{2l}{c}\approx 13.3ns$ is the cavity round-trip time. The $200\mu m$ diameter of each site in the array ensures a single Gaussian spatial mode for each laser.  

Intra-cavity polarizing beam splitter (PBS) and $\lambda/2$-waveplate deflect a controllable amount of the light into a second branch of the cavity \cite{cohen2009single,fridman2010phase,smith1965stabilized} where the lasers are Talbot-coupled \cite{tradonsky2017talbot,talbotLeger}. 
This provides a tunable coupling strength of $K_{max}\sin[2](2\theta)$ between nearest neighbor lasers \cite{Pando:23},
where 
 $\theta$ is the rotation angle of the $\lambda/2$-waveplate and
$K_{max}\approx -0.45$ is the calculated full Talbot coupling strength \cite{tradonsky2017talbot,talbotLeger}. 

In each experimental realization, the SLM was controlled to obtain a normally distributed random frequency detuning pattern with a standard deviation $\Omega_{RMS}$ and a Gaussian spatial correlation function with a waist that we refer to  as $\xi$, the correlation length of the frequency detuning pattern. Hence, the correlation of the frequency detuning between the lasers in sites  $(i,j)$ and $(i',j')$ is:
\begin{equation}\label{eqn_xi}
    \ev{\Omega_{ij}\Omega_{i'j'}} = (\Omega_{RMS})^2e^{-\frac{(i-i')^2+(j-j')^2}{\xi^2}}
\end{equation}

Eq. (\ref{eqn_xi}) indicates that as $\xi$ increases, the detunings of neighboring lasers are more likely to be similar.
To provide near-perfect starting conditions for our experiments we first apply intra-cavity adaptive optics \cite{Pando:23} to reduce aberrations and uncontrolled frequency detuning (see below and Supplemental Material \cite{Supplemental}, Fig. S2). 

For each realization, we pump the laser and measure the resulting steady state near field (NF) and far field (FF) intensity distributions. The measured distributions are averaged over 50 random realizations for each value of $\tau_c\Omega_{RMS}$ and $\xi$. The inital pump power was $P=19.8W \approx 4\cdot P_{th}$, and increased as required to compensate for the increased losses due to the introduced disorder (see Fig. S1 in \cite{Supplemental}). The FF intensity distribution ($I_{FF}$) is proportional to the Fourier transform of the coherence function of the electric field \cite{mandel1965coherence,friberg1983spatial}. We thus use the average FF inverse participation ratio (IPR) as the synchronization order parameter:
\begin{equation}\label{eqn_IPR}
    \text{IPR}=\frac{\int\int\dd{x}\dd{y}I_{FF}^2}{(\int\int\dd{x}\dd{y}I_{FF})^2}.
\end{equation}
The IPR is a common measure of localization in distributions and is correlated with the average number of synchronized (i.e. mutually coherent) lasers (see Fig. S4 in \cite{Supplemental}) \cite{Marc_ipr,longhi2022non, Pando:23,friberg1983spatial}. For an array of lasers, each with a single Gaussian mode, and with a mutual coherence length $l_c$, $\text{IPR}\propto l_c^2\equiv A_c$. We interpret $A_c$ as a coherence area and $N = A_c/d_{lat}^2$ as the average number of synchronized (mutually coherent) lasers\cite{Supplemental}. The proportionality constants are determined by the geometry of the system, and remain constant throughout the experiments. In earlier investigations we found that the results of the IPR measurements of the laser array coherence were equivalent to interferometric phase measurements or spectral frequency measurements, while being simpler and more reliable for large and disordered arrays \cite{pal2020rapid,fridman2010phase,chriki2018spatiotemporal}.

\begin{figure}[h]
    \centering
    \includegraphics[trim=16cm  0cm 16.5cm 0cm,clip,width=\linewidth]{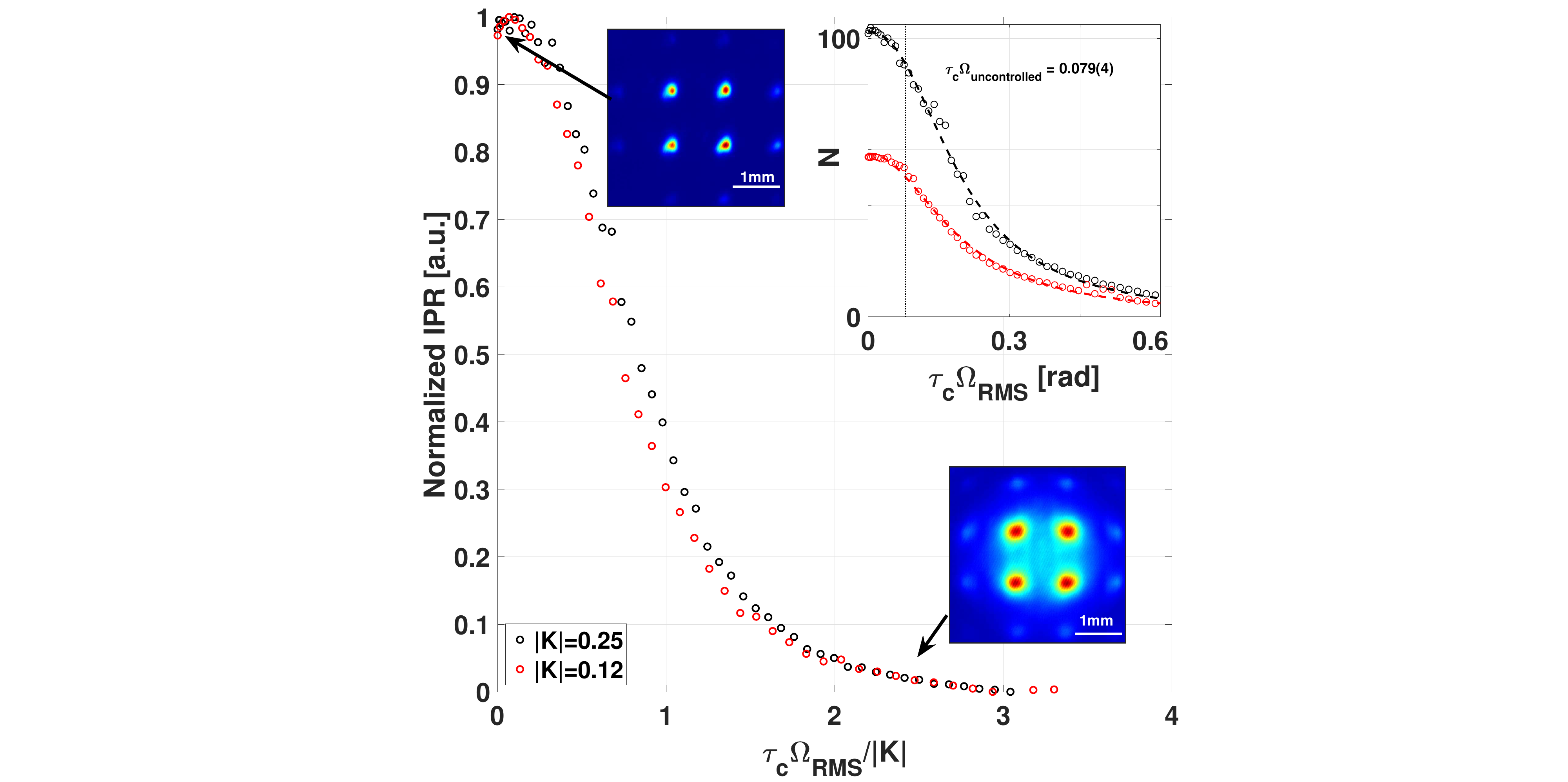}
    \caption{Experimental normalized FF IPR as a function of the ratio of disorder over the coupling strength, $\tau_c\Omega_{RMS}/\abs{K}$ for two coupling values. Insets show the average FF intensity distribution at $\tau_c\Omega_{RMS}/\abs{K}=0 \text{ and }2.48$ for coupling strength $\abs{K}=0.25$. \textbf{Top right inset:} Average number of synchronized lasers , $N\equiv A_c/d_{lat}^2$, as a function of disorder. Dashed curves denote best fits to $N=\frac{a}{(\tau_c\Omega_{RMS})^b+c}$, and the dotted vertical line denotes the estimated value of $\tau_c\Omega_{\text{RMS, uncontrolled}}=0.079 \text{ rad}$. }
    \label{fig_coupling}
\end{figure}

\textit{Results, uncorrelated disorder - } Figure  \ref{fig_coupling} shows the experimental normalized FF IPR as a function of the applied disorder strength (normalized to the coupling strength) $\tau_c\Omega_{RMS}/\abs{K}$, for uncorrelated ($\xi=0$) normally distributed frequency detuning patterns. As evident, increasing the disorder leads to a monotonic decrease in the IPR and deterioration of synchronization, as manifested by the significant broadening of the FF intensity peaks, shown in the insets (see also Fig. S4 in \cite{Supplemental}). Results for different $\abs{K}$ agree well with each other, with the IPR dropping to half for $\tau_c\Omega_{RMS}/\abs{K}\approx 0.81(7)$, attesting that synchronization is determined by the ratio $\tau_c\Omega_{RMS}/\abs{K}$.

The top right inset in Fig. \ref{fig_coupling} shows that the average number of synchronized lasers (determined from the FF intensity distribution \cite{Pando:23,friberg1983spatial}) is also reduced monotonically with  $\tau_c\Omega_{RMS}$. Note that $N$ differs significantly for the two different coupling strengths for small $\tau_c\Omega_{RMS}$ due to the uncontrolled disorder in our system. 

Fitting the data from Fig.$\,\,\,$\ref{fig_coupling} top right inset to $N=\frac{a}{(\tau_c\Omega_{RMS})^b+c}$ we obtain $b=2.3(1),2.0(1)$ for $\abs{K}=0.25, 0.12$, respectively. These results are in good agreement with the theoretical value of $b=2$ in Eq.(\ref{eqn_N_prop}) which we consider in the last section of this paper. We  identify $c$ as a manifestation of our uncontrolled disorder $(\tau_c\Omega_{\text{RMS, uncontrolled}})^b = c$, to estimate  $\tau_c\Omega_{\text{RMS,uncontrolled}} \approx 0.079(4) \text{ rad}$.

\textit{Results, correlated disorder - } We now consider the effects of correlated disorder ($\xi>0$) on synchronization. The results for $\abs{K}=0.25$ are presented in Figs. \ref{fig_phasediag}-\ref{fig_LRE_xi} (see additional results in Figs. S5-S7 in \cite{Supplemental}). Figure \ref{fig_phasediag} shows the experimental normalized IPR of the measured FF intensity distribution as a function of $\tau_c\Omega_{RMS}$ and $\xi$. All measured IPR values were normalized such that $\text{IPR} = 1,0$ are the maximal and minimal values measured across all experiments, respectively. For all values of $\xi$, the IPR monotonically decreases as $\tau_c\Omega_{RMS}$ is increased, as expected. However, the IPR dependence on $\xi$ is non-trivial and non-monotonic.
\begin{figure}[h]
    \centering
    \includegraphics[trim=18cm  0cm 18cm 0.5cm,clip,width=\linewidth]{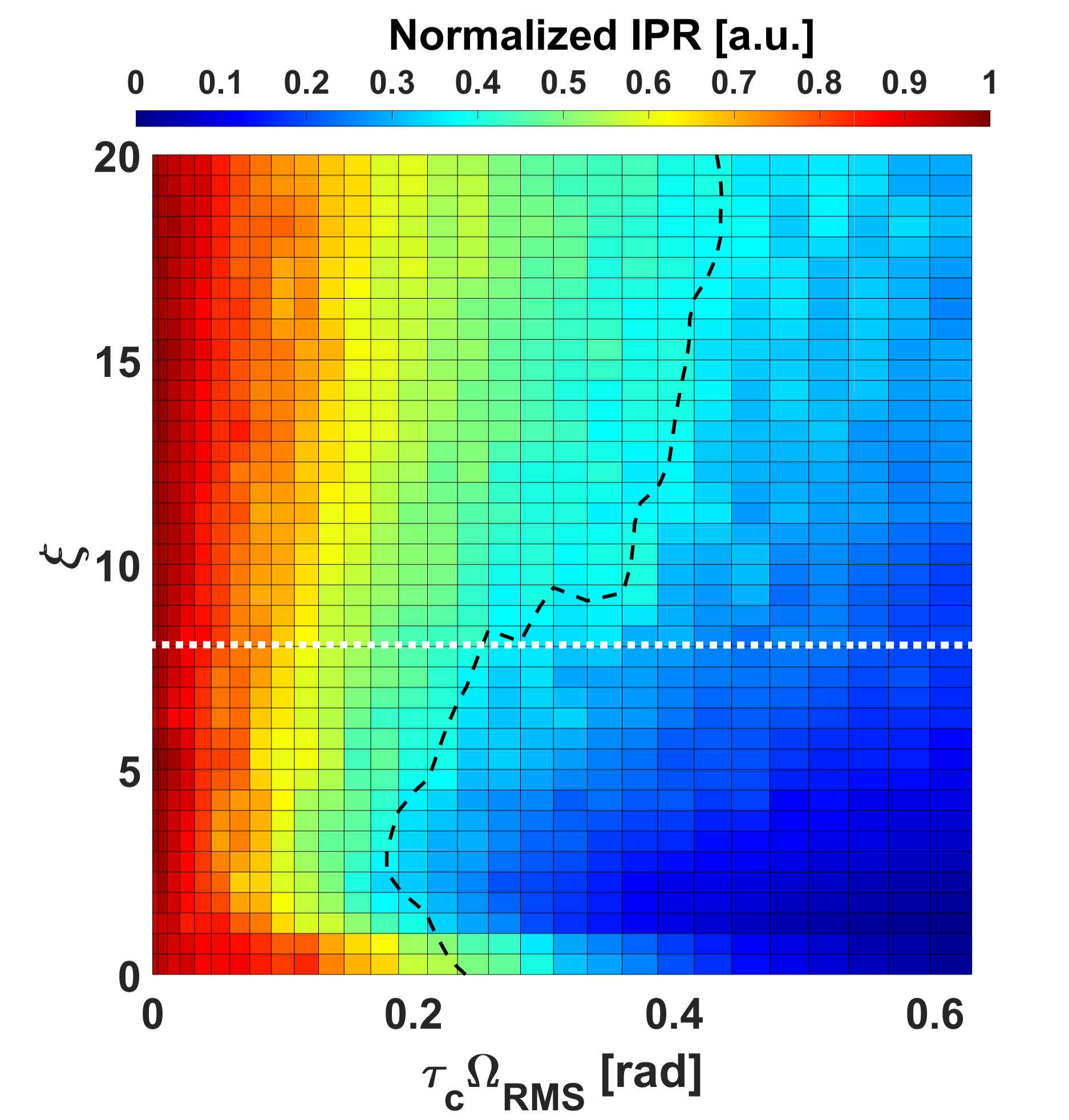}
    \caption{Measured FF IPR as a function of the disorder strength $\tau_c\Omega_{RMS}$ and its correlation length $\xi$ for $\abs{K}=0.25$. The black dashed curve denotes $\text{IPR}=0.4$, indicating the non-monotonic dependence on $\xi$. The white dotted line is along $\xi=8$ (detailed in Fig. \ref{fig_LRE_det}). }
    \label{fig_phasediag}
\end{figure}

Figure \ref{fig_LRE_det} shows the experimental normalized and calculated IPR as a function of  $\tau_c\Omega_{RMS}$ for uncorrelated disorder ($\xi=0$) and correlated disorder with $\xi=8$. For weak disorder $\tau_c\Omega_{RMS}<0.31\text{ rad}$, the IPR is lower (worse synchronization) for the correlated disorder, while for strong disorder, the opposite is true. Numerical simulation of the full laser rate equations \cite{PhysRevLett.92.093905}( see procedure in \cite{Supplemental}) are in good agreement with the experimental results and validate the non-monotonic dependence on $\xi$.

This non-monotonicity is again seen in Fig.$\;$\ref{fig_LRE_xi} that shows the experimental and calculated normalized IPR as a function of $\xi$ for different values of $\tau_c\Omega_{RMS}$. In the case of strong disorder (purple), the IPR monotonically increases with $\xi$. However, for weak disorders the dependence on $\xi$ is non-monotonic, and the IPR reaches a minimum at an intermediate value of $\xi$ that decreases with the strength of the disorder. It is apparent that there is a good agreement between the experimental and simulation results.
\begin{figure}[h]
    \centering
    \includegraphics[trim=1cm  0.5cm 1cm 2.5cm,clip,width=1.1\linewidth]{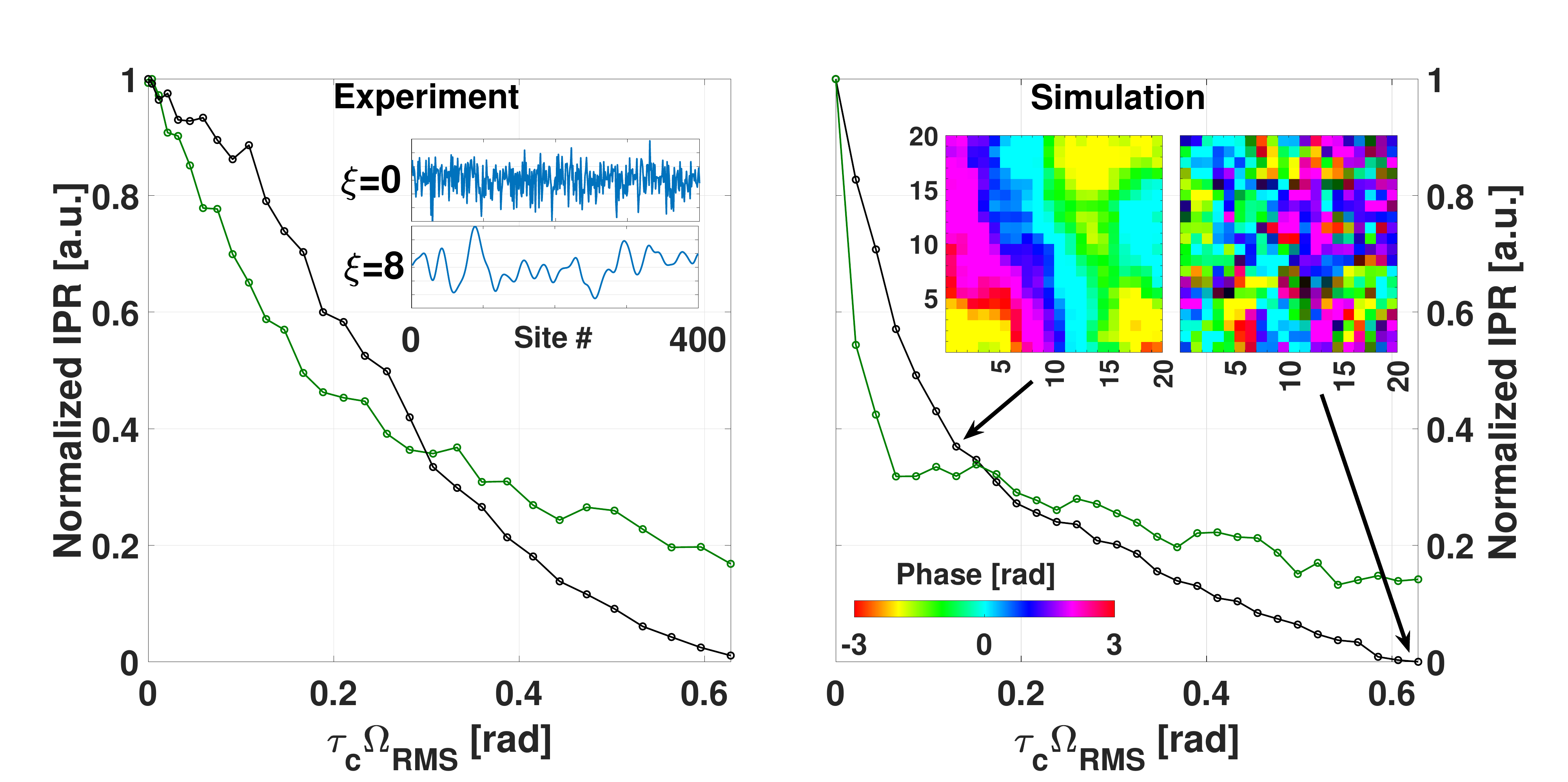}
    \caption{Experimental and numerically simulated normalized FF IPR as a function of disorder $\tau_c\Omega_{RMS}$ for uncorrelated disorder $\xi=0$ (black) and for correlated disorder with $\xi=8$ (green). \textbf{Inset, left:} Representative realizations of disorder vectors with the same length (400 sites) and $\tau_c\Omega_{RMS}$ for $\xi=0$ and $\xi=8$. \textbf{Inset, right:} Examples of synchronized lasers in LRE simulations for uncorrelated disorder ($\xi=0$) with $\tau_c\Omega_{RMS} = 0.15, 0.62\; \text{rad}$. Each pixel represents the phase of the corresponding laser in the array. Additional examples are provided in Fig. S3 in \cite{Supplemental}.}
    \label{fig_LRE_det}
    \hfill \break
    \centering
    \includegraphics[trim=0cm  1cm 0cm 2cm,clip,width=1.1\linewidth]{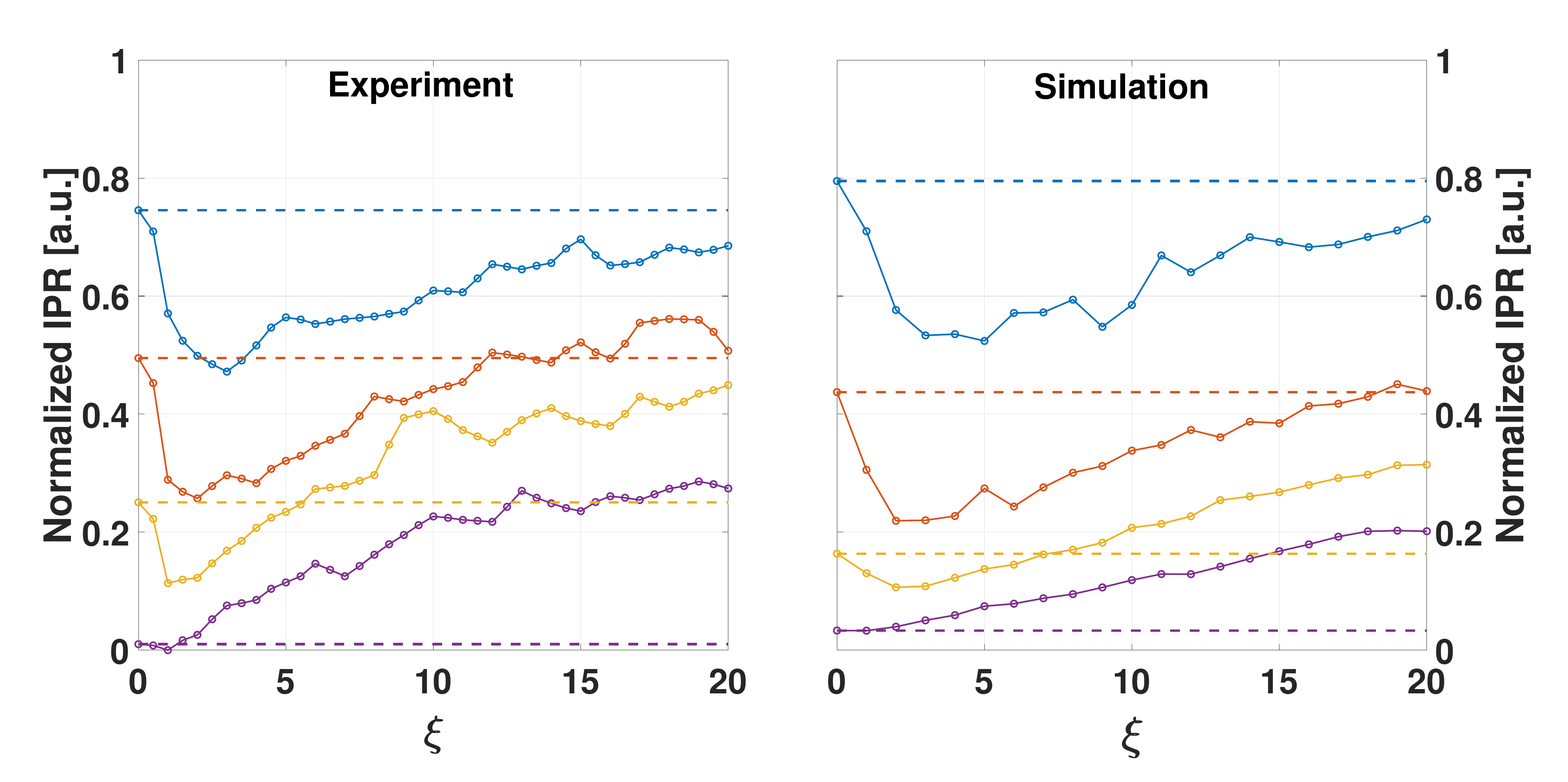}
    \caption{Experimental and numerically simulated normalized FF IPR as a function of correlation length $\xi$ for different values of disorder, $\tau_c\Omega_{RMS}=0.12, 0.23, 0.36, 0.62 \text{ rad}$ (blue, orange, yellow and purple respectively). Dashed lines show the corresponding IPR values for $\xi=0$ for reference.}
    \label{fig_LRE_xi}
\end{figure}

\textit{Analysis and discussion - }
To elucidate our experimental and simulation results, we consider a toy model that is based on the Kuramoto model\cite{ acebron2005kuramoto}, which can describe the dynamics of coupled lasers when their intensities are identical and high above threshold \cite{nixon2013, honari2020mapping}. 
Theoretical studies of models with nearest neighbor coupling have shown that for any finite frequency detuning $\Omega_i$ (taking $\tau_c=1$), the oscillators synchronize in local clusters, where the maximal number of synchronized oscillators in a single cluster is bounded \cite{acebron2005kuramoto,strogatz1988collective,sakaguchi1987local}.

 Specifically, we consider a one-dimensional chain of Kuramoto oscillators with bi-directional nearest neighbor coupling as a toy model of our system. We chose to work with a one-dimensional theoretical system for which analytical solutions exist (rather than the two-dimensional system in our experiments and simulations where analytical solutions do not exist). Although it is not obvious that results from a one-dimensional system can be applied to a two-dimensional one, theoretical studies that suggest that the conditions for existence of the phase locked state of the system and its’ properties are similar for one and two-dimensional systems \cite{sakaguchi1987local,strogatz1988collective}. We found that our limited toy model provides some insight into the results of our two-dimensional system, and shows similar qualitative behavior to that which we have observed in experiments.

For a one-dimensional chain of oscillators with nearest neighbor coupling, the necessary condition for synchronization of $N$ oscillators is that the maximal accumulated detuning along the $N$ oscillators must be smaller than the coupling strength $K$ between two neighbors \cite{strogatz1988collective}:
\begin{equation}
    \label{eqn_strogatz}
    \max_{1\le j\le N}{\abs{X_j}}\le K
\end{equation} 
with $X_j$ being the accumulated detuning:
\vspace{-3pt}
\begin{equation}
    \label{xj_strogatz}
    X_j = \sum_{i=1}^j\Omega_i - \frac{1}{N}(\sum_{i=1}^N\Omega_i).
\end{equation}

$K_c$, the critical required coupling for synchronization of $N$ oscillators is thus simply  $\max{\abs{X_j}}$. When $\Omega_j$ has a normal distribution (i.e., for uncorrelated disorder), Eq. (\ref{eqn_strogatz}) describes the maximal displacement of a random walker, $\max{\abs{X_j}}\propto \Omega_{RMS}\sqrt{N}$, such that the synchronized cluster size is 
\begin{equation}
    \label{eqn_N_prop}
    N\propto \frac{K^2}{\Omega_{RMS}^2}. 
    \end{equation} 
Eq. (\ref{eqn_N_prop}) agrees well with the results presented in Fig. \ref{fig_coupling}, as well as the fit to the experimental data.

\begin{figure}[h]
    \vspace{-2mm}
    \centering
    \includegraphics[trim=5cm 0.5cm 0.5cm 0cm,clip,width=1.1\linewidth]{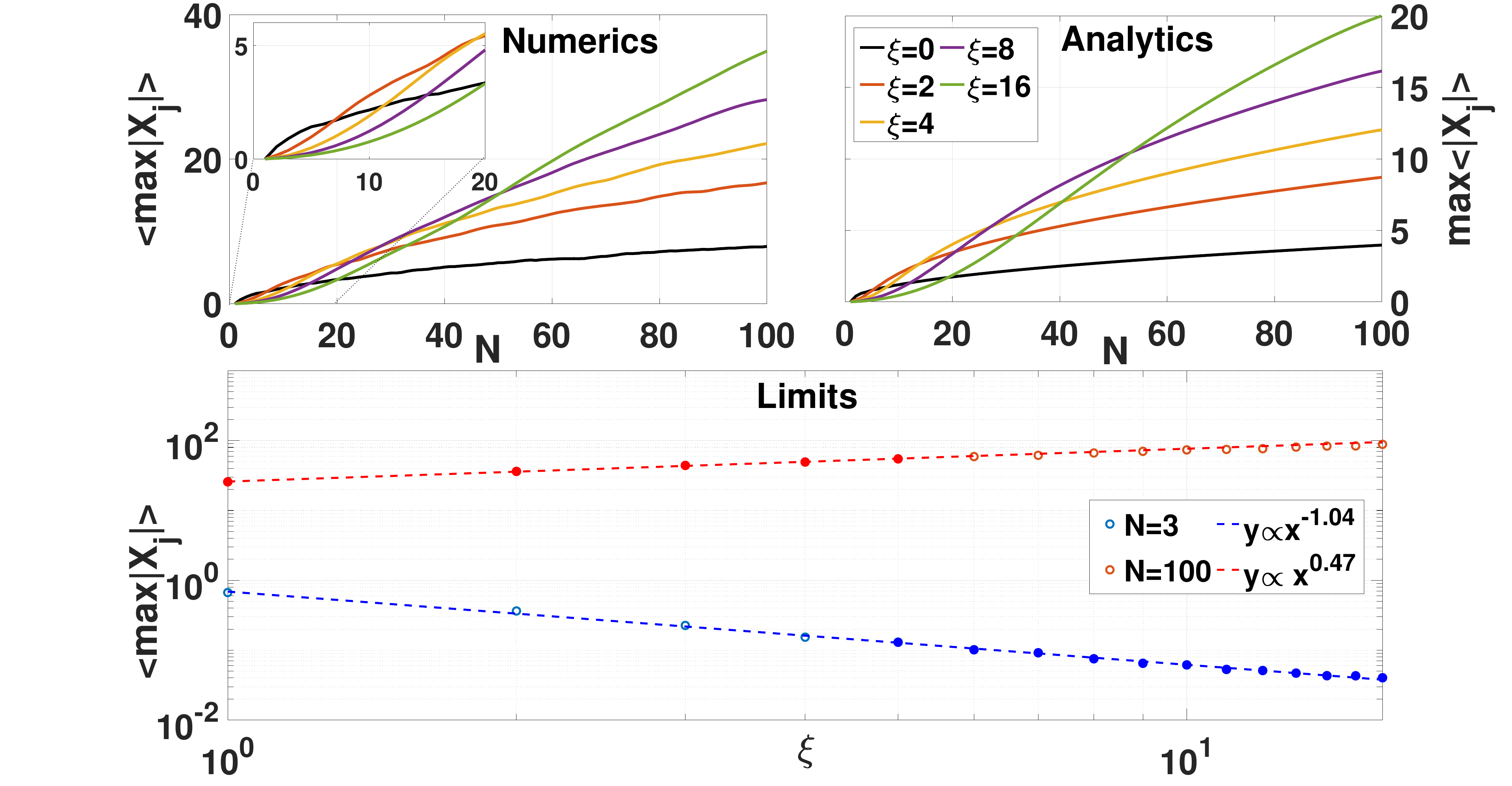}
    \caption{The maximal accumulated detuning $\max{\ev{\abs{X_j}}}$ as a function of the number of oscillators N and disorder correlation length $\xi$. \textbf{Upper left}: Numerical integration of Eq. (\ref{xj_strogatz}) averaged over 100 random realizations. The inset shows a magnified view for low N values. 
    \textbf{Upper right}: Corresponding analytical evaluation using Eq.(\ref{eqn_analytic}). 
    \textbf{Bottom}: Log-log plot of the numerical data as a function of $\xi$ for $N=100$ (red) and $N=3$ (blue) $N$. Linear fits to the colored-in points (dashed lines) yield good agreement of the $\xi$ scaling to the limiting analytical approximations.}
    \label{fig_numerical_compare}
    \vspace{-4mm}
\end{figure}

We now extend our model for the case of correlated quenched disorder. Figure \ref{fig_numerical_compare} (upper left) shows  $\ev{\max{\abs{X_j}}}=K_c$ numerically calculated from Eq. (\ref{xj_strogatz}) as a function of the synchronized cluster size $N$ for Gaussian detuning disorder of $\Omega_{RMS}=1$ and several correlation lengths $\xi$ (the brackets indicate an average over different disorder realizations).
For small synchronized cluster sizes $K_c$ decreases with $\xi$, while for larger cluster sizes, $K_c$ increases with $\xi$. Notably, the crossing point between the two trends is roughly at $N\sim \xi$. 
 
The result of Eq. (\ref{xj_strogatz}) in the case of correlated disorder can be approximated analytically (see derivation in \cite{Supplemental}) as:  
\begin{equation}
  \begin{aligned}
    \ev{\abs{X_j}}^2 & \approx \Omega_{RMS}^2\frac{\pi \xi L}{8}\frac{1}{\frac{\xi}{\sqrt{2\pi}}(e^{-\frac{2L^2}{\xi^2}}-1)+L\operatorname{erf}(\frac{\sqrt{2}L}{\xi})}\times  \\
   \sum_{i=0}^L &  [\operatorname{erf}(\frac{j-i}{\xi})+\frac{1}{2}e^{-\frac{(j-i)^2}{\xi^2}}-\frac{j}{N}(\operatorname{erf}(\frac{N-i}{\xi})\\ 
   & +\frac{1}{2}e^{-\frac{(N-i)^2}{\xi^2}})+(1-\frac{j}{N})(\operatorname{erf}(\frac{i}{\xi})-\frac{1}{2}e^{-\frac{i^2}{\xi^2}}) ]^2
  \end{aligned}
  \label{eqn_analytic}
\end{equation}
The analytic approximations of Eq.(\ref{eqn_analytic}) shown in Fig. \ref{fig_numerical_compare} (upper right) are in good agreement with the exact numerical integration of Eq. (\ref{xj_strogatz}) (Fig.  \ref{fig_numerical_compare} upper left) and validate the non-monotonic dependence of $K_c$ on $\xi$ and the cluster size. 

Analyzing the limiting behavior of Eq.(\ref{eqn_analytic}) reveals two distinct regimes. In one regime where $N\gg \xi$, $\max\ev{\abs{X_j}}\rightarrow \sqrt{\xi N}$, equivalent to the displacement of a random walker with a step size $\xi>1$. In the other regime, $N\ll \xi$, $\max\ev{\abs{X_j}}\rightarrow \frac{N^2}{\xi}$, which can also be derived directly from Eq.(\ref{eqn_strogatz}) by treating the applied disorder as a long wavelength perturbation $\Omega_i=\sin(\frac{i}{\xi})$. A log-log linear fit to the numerically evaluated $\ev{\max\abs{X_j}}$ as a function of $\xi$ is shown in Fig. \ref{fig_numerical_compare} (bottom) for $N=3$ and $N=100$. Both regimes are well fitted by $y=ax^b$ with $b=-1.04(6), 0.47(4)$ for the $\xi\gg N,\xi\ll N$ regimes, in good agreement with the theoretical limiting behavior of $b=-1, 0.5$, respectively. The results from the analytical toy model reveal a behavior which is qualitatively similar to the non monotonic relationship between the disorder parameters $\tau_c\Omega_{RMS},\xi$ and the synchronization of the array shown in Figs. \ref{fig_phasediag}-\ref{fig_LRE_xi}.

\textit{Conclusions - } We experimentally investigated the effects of quenched disorder on the synchronization of coupled oscillators by means of frequency detuning disorder in coupled laser arrays. 
Our results demonstrate how increased disorder results in a gradual deterioration in synchronization that depends on the ratio of the coupling strength over the disorder strength. Our experimental results are supported by both numerical simulations and an analytic toy model. In addition, we found that the correlated disorder can either improve or degrade synchronization compared to uncorrelated disorder, depending on the ratio of its correlation length $\xi$ and the average number of synchronized lasers $N$: For $\xi\ll N$, $N\propto \frac{K^2}{\xi^2\Omega_{RMS}^2}$ revealing the behavior of a correlated random walker. In contrast, disorder with $\xi\gg N$ is effectively a low frequency perturbation along the cluster and thus causes a smaller decay in synchronization to yield $N\propto \sqrt{\frac{\xi K}{\Omega_{RMS}}}$. 

Our results provide insight into the effects and management of disorder which can be exploited to improve systems where disorder has an inherent correlation time or length (e.g. spin and photonic systems). By controlling the applied disorder, it should be possible to quantify protection against disorder by means of topological effects \cite{alex2021,bandres2018topological,rosiek2023observation,wang2009observation,hafezi2011robust,contractor2022scalable}, and study the effects of disorder on spin simulators and solvers which are based on coupled lasers or parametric oscillators \cite{wang2013coherent,marandi2014network,mcmahon2016fully}.

\FloatBarrier

\clearpage
\newpage
\widetext
\begin{center}
\textbf{\Large Supplemental Materials}
\end{center}
\setcounter{figure}{0}
\renewcommand{\figurename}{Fig.}
\renewcommand{\thefigure}{S\arabic{figure}}

\section{Maintaining near field intensity distribution}

\begin{figure}[H]
    \centering
    \includegraphics[trim=15cm 1cm 15cm 1cm, clip,width=0.5\linewidth]{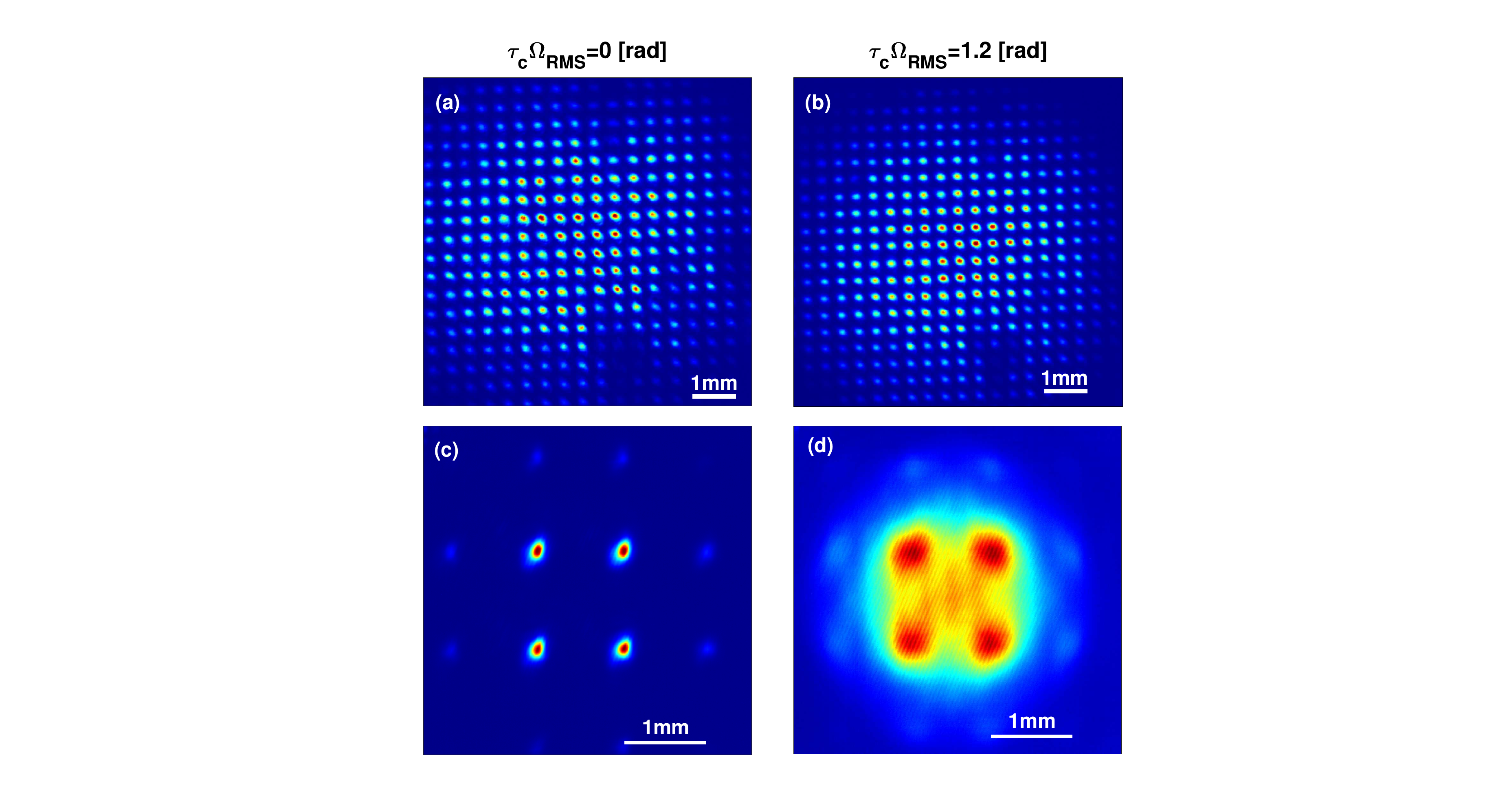}
    \caption{Normalized average NF and FF intensity distributions for $\tau_c\Omega_{RMS}=0\text{rad}$ \textbf{(a),(c)} and for $\tau_c\Omega_{RMS}=1.2\text{ rad}$\textbf{(b),(d)}.}
    \label{fig_nfff}
\end{figure}

As noted in the main text, the coupled lasers array can be described by the Kuramoto model equations when the lasers' intensities are equal. However, the introduction of frequency detuning to the lasers also induces increased losses, caused by the decrease in phase synchronization (i.e. phase locking or mutual coherence) and partial coupling between loss and detuning in our system with the SLM. In order to avoid lasing non-uniformities and remain high above the lasing threshold of the entire array, we gradually varied our pump power so as to maintain roughly equal NF intensity. Specifically, for each value of $\tau_c\Omega_{RMS}$ we measured the average total NF intensity, $I_{NF}^{tot}$, and compared it to its initial value at $\tau_c\Omega_{RMS}=0$. We increased or decreased the pump power so that $I_{NF}^{tot}$ would be within $5\%$ of its initial value. Figure \ref{fig_nfff} shows the NF and FF intensity distributions at the minimal and largest $\tau_c\Omega_{RMS}$ applied in the experiment. It can be seen that the NF intensity distribution is essentially not affected by the increased disorder, while the FF intensity distribution shows vastly reduced phase synchronization, as indicated by the broad diffraction peaks and strong background.
\newpage
\section{The effect of intra-cavity adaptive optics}
\begin{figure}[H]
    \centering
    \includegraphics[trim=7cm  0cm 7cm 1cm,clip,width=0.5\linewidth]{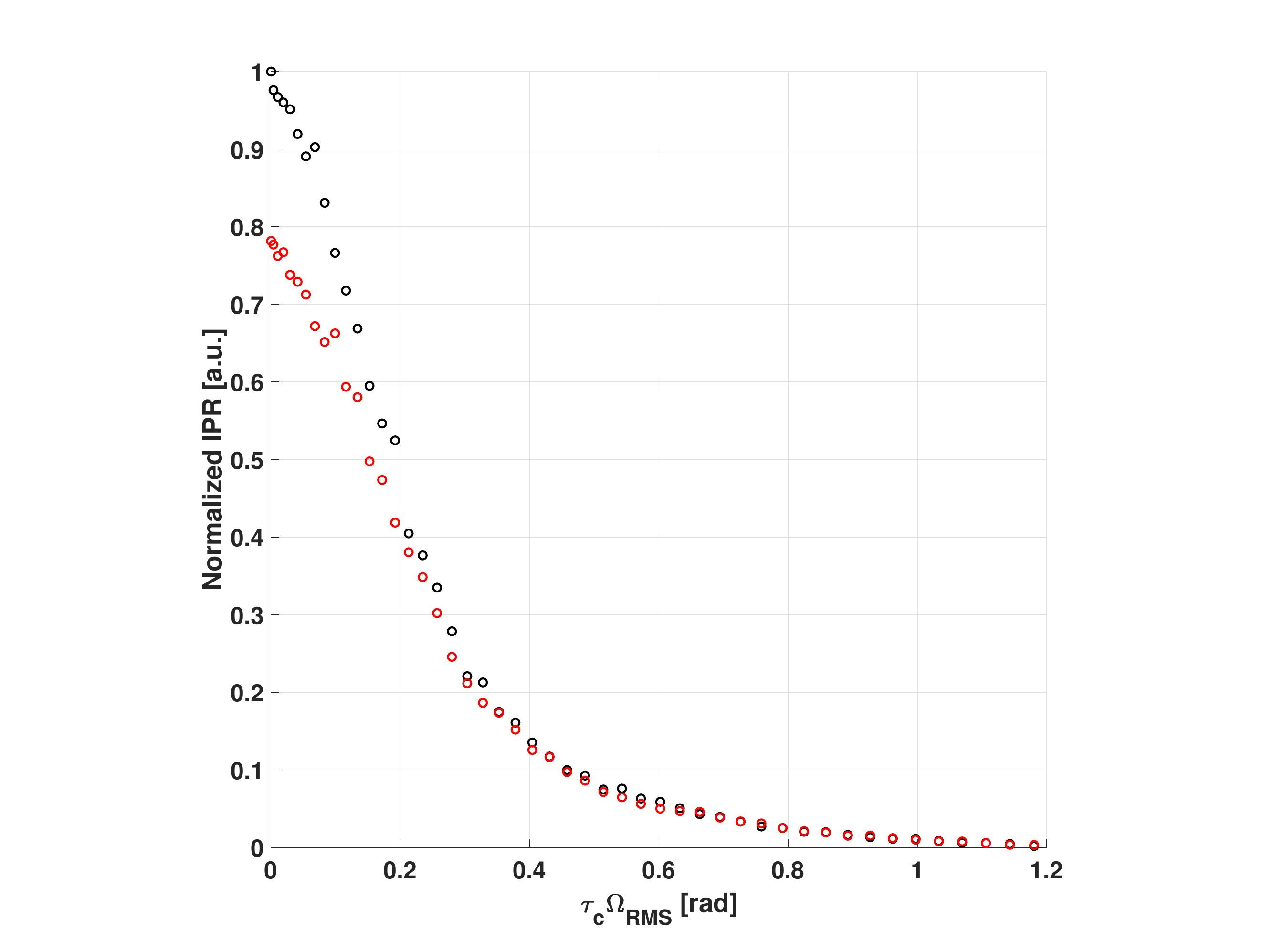}
    \caption{Experimentally measured normalized FF IPR with (black) and without (red) adaptive optics correction. }
    \label{fig_AO}
\end{figure}

In an earlier work, we developed an intra-cavity adaptive optics method (AO) and demonstrated its beneficial effect on phase synchronization\cite{Pando:23}. We resorted to the same method in order to reduce the amount of uncontrolled detuning disorder in the cavity. Figure \ref{fig_AO} shows the experimental normalized FF IPR as a function of disorder with and without the application of adaptive optics. The results show that initial phase synchronization (at $\Omega_{RMS}=0$) is improved by approximately $25\%$ as measured by the IPR. Furthermore, for large $\tau_c\Omega_{RMS}$ values, the measurements coincide. This is likely due to the fact that for these values, $\tau_c\Omega_{\text{RMS, uncontrolled}}\ll\tau_c\Omega_{RMS}$, and hence the quality of phase synchronization is determined only by the change in $\tau_c\Omega_{RMS}$. We note that this happens at a relatively large value of $\tau_c\Omega_{RMS}$, hence highlighting the importance of using our AO method to optimize cavity performance.
\newpage
\section{Numerical simulation of the full Laser Rate Equations (LRE)}

In addition to experiments, we study the system numerically using the Laser Rate Equations (LRE)\cite{PhysRevLett.92.093905}, a set of coupled differential equations which simulate the dynamics of a coupled laser system. The LRE are given by:
\begin{equation} \tag{E1}
\dv{E_m}{\tau_c}=\frac{G_m-\alpha_m+i\tau_c\Omega_m}{\tau_c}E_m+\frac{1}{\tau_c}K_{mn}E_n
\end{equation}
$$\dv{G_m}{\tau_c}=\frac{1}{\tau_f}[P_m-G_m(\frac{\abs{E_m}^2}{I_{sat}}+1)]$$

Where $E_m,P_m,G_m,\alpha_m,\Omega_m$ are the (complex) electric field, pump power, gain, loss, and detuning of the m-th laser, respectively. $\tau_f$ and $\tau_c$ are the fluorescence lifetime and cavity round trip time, and $I_{sat}$ is the saturation intensity of the gain. Finally, $K_{mn}$ is the coupling term between the m-th and n-th laser. By changing the value of $K_{mn}$ we can vary the coupling scheme of the simulation to represent different coupling ranges and geometries. In all of the simulations mentioned in this work, we use a nearest neighbor coupling scheme over a square lattice, such that each laser is coupled to its four neighbors.

Denoting $E_m=A_m e^{i\phi_m}$, we can divide Eq. (E1) its real and imaginary components, affecting the amplitude and phase of the electric field respectively:
\begin{equation}\tag{E2}
    \dv{A_m}{\tau_c} = \frac{G_m-\alpha_m}{\tau_c}A_m+\frac{1}{\tau_c}K_{mn}A_n\cos(\phi_m-\phi_n) 
\end{equation}
\begin{equation}\tag{E3}
    \dv{\phi_m}{\tau_c} = \Omega_m + \frac{1}{\tau_c}K_{mn}\frac{A_n}{A_m}\sin(\phi_m-\phi_n)
\end{equation}

In the limit where the lasers' amplitudes are equal, $A_m=A$, and taking $\tau_c\rightarrow 0$, Eq. (E3) can be identified as the Kuramoto model equation,
$$\dot{\phi_m} = \Omega_m + K_{mn}\sin(\phi_m-\phi_n)$$
up to a rescaling of the coupling and detuning terms. Hence, in the case of equal laser intensities, we expect Kuramoto-like dynamics to emerge from the LRE.

For the numerical results presented in the paper, we solved the LRE for an array of $20\times 20$ lasers coupled with bi-directional nearest neighbor negative coupling. The parameters that we used were similar to the experimental parameters, with the loss set to $\alpha_0 = 1.7$ and the nearest neighbor coupling set to $K= -0.25$. The simulation starts from a cold cavity, with the gain $G=0$ for all lasers, and the pump power $P=4 \cdot P_{th}$, where $P_{th}$ is the pump threshold value. The pump power was chosen to fit the experimental parameters, and in a similar fashion to the experiment, it was modified to maintain equal average lasers intensities throughout (as noted in section I of the Supplemental Material above).
in S1). We additionally set the fluorescence lifetime $\tau_f=10^3 \tau_c$, similar to the actual experimental value.
In each iteration of the simulation, we chose a value of the correlation length $\xi \in [0,20]$ and disorder $\tau_c\Omega_{RMS}\in [0, 0.62]\text{ rad}$. We generate a random detuning matrix $\tau_c\Omega_{ij}$, with $i,j=1,...20$ in a similar fashion to the experiment, such that it has a standard deviation $\tau_c\Omega_{RMS}$ and correlation length $\xi$. Each point was repeated for 50 different realizations, and the results were averaged accordingly. The initial conditions for each realization were also randomly generated. The simulation involved $10^5$ iterations, similar to the duration of the quasi-CW pulses of our experimental system.

The results were Fourier transformed to obtain far field intensity distributions, similar to those observed experimentally. We then calculated the FF IPR, and the results are presented in Figs.4-5 in the main text.
\newpage
\section{Visualizing Synchronized Clusters}
\begin{figure}[H]
    \centering
    \includegraphics[width=\linewidth]{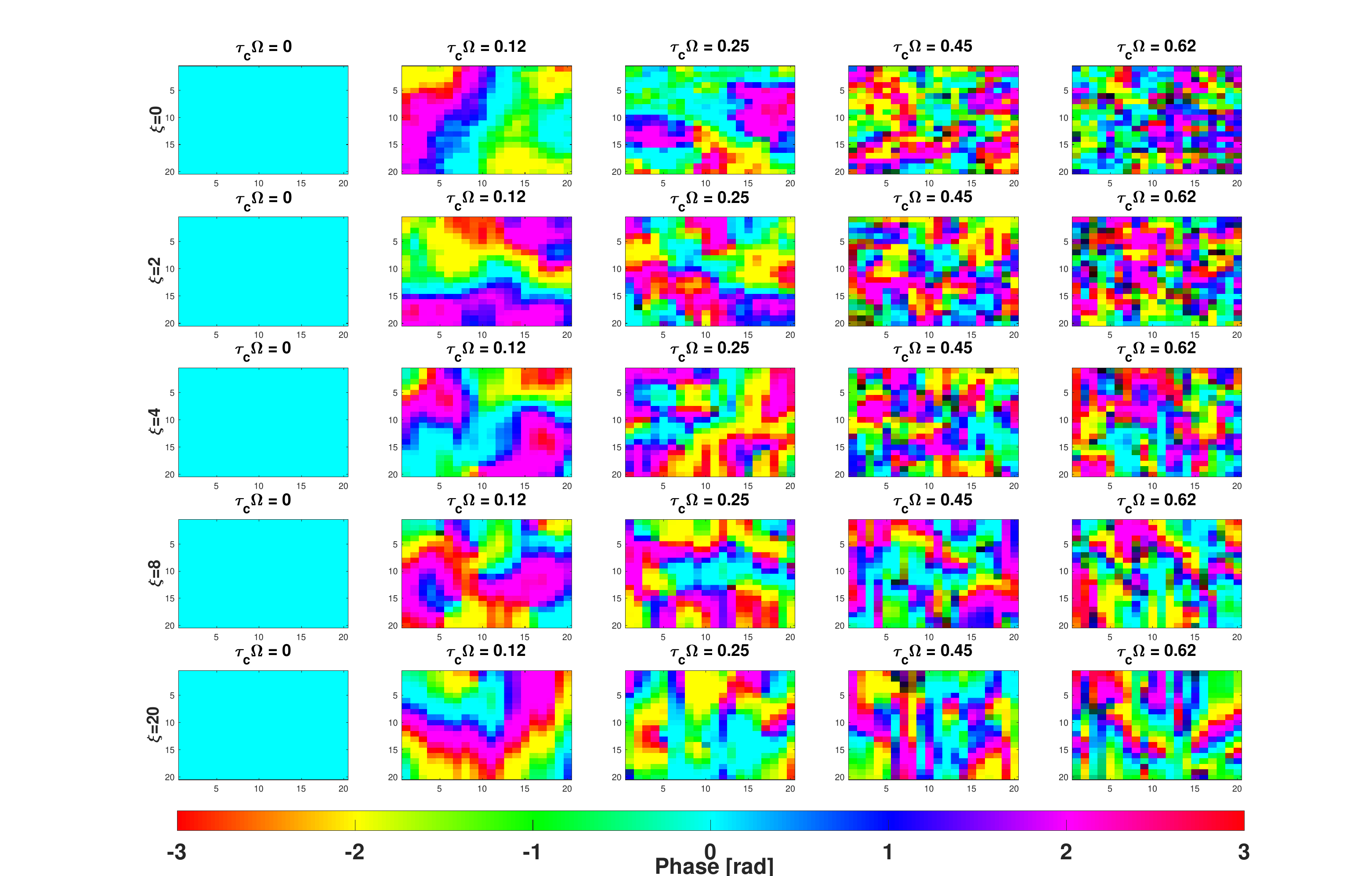}
    \caption{Illustrations of synchronized phase clusters in example realizations of LRE simulations. Each pixel in the $20\times 20$ grid  the phase of the corresponding laser. Different columns correspond to different disorder strengths $\tau_c\Omega_{RMS}$ in $rad$, and different rows correspond to different choices of disorder correlation length $\xi$.}
    \label{fig_phase_example}
\end{figure}

Previous works on locally coupled Kuramoto oscillators have shown that the oscillators split into locally separate synchronized clusters \cite{strogatz1988collective,sakaguchi1987local}. We performed numerical simulations (using LRE) of a coupled $20\times20$ laser array where the frequency detuning randomly varied. The simulation used positive coupling for ease of visualization. Figure \ref{fig_phase_example} shows example realizations of the laser phases at the end of these simulations, for different magnitudes of disorder and correlation lengths. 
\newpage
\section{Quantifying phase synchronization with IPR}
As mentioned in the main text, there is a relation between the coherence length of the laser array, $l_c$, and the far field intensity distribution \cite{friberg1983spatial,mandel1965coherence}. For example, consider a Gaussian beam with waist $w$ in the transverse axes and some finite spatial coherence length. The transverse coherence length of the beam, $l_c$ is inversely proportional to its far field diffraction angle\cite{friberg1983spatial}, and proportional to $w$. The IPR of a the same Gaussian beam is 
$$\text{IPR}_{\text{Gaussian}} \propto w^2 \propto l_c^2 \equiv A_c,$$
which can be interpreted as the coherence area of the beam.  We extend this result for an array of Gaussian beams, such as our laser array. Each indivdual beam in the array has waist $w$ and the distance between two neighboring beams is $d_{lat}$. We also assume that the mututal coherence between beams in the array decays with a Gaussian enevlope with scale $l_c$. Assuming each individual Gaussian is a coherent source, the IPR yields:
$$\text{IPR}_{\text{array}} = C(w,d_{lat})l_c^2 $$
where $C$ is a prefactor set by the shape of the array - the waist of each individual laser and the spacing between them, $d_{lat}$. This result is important as it shows that the IPR allows us to take complicated spatial functions, namely the far field intensity distribution of the field and the second order coherence function, and condense them to a single meaningful number. 

\begin{figure}[H]
    \centering
    \includegraphics[trim=2cm  0cm 4cm 1cm,clip,width=0.7\linewidth]{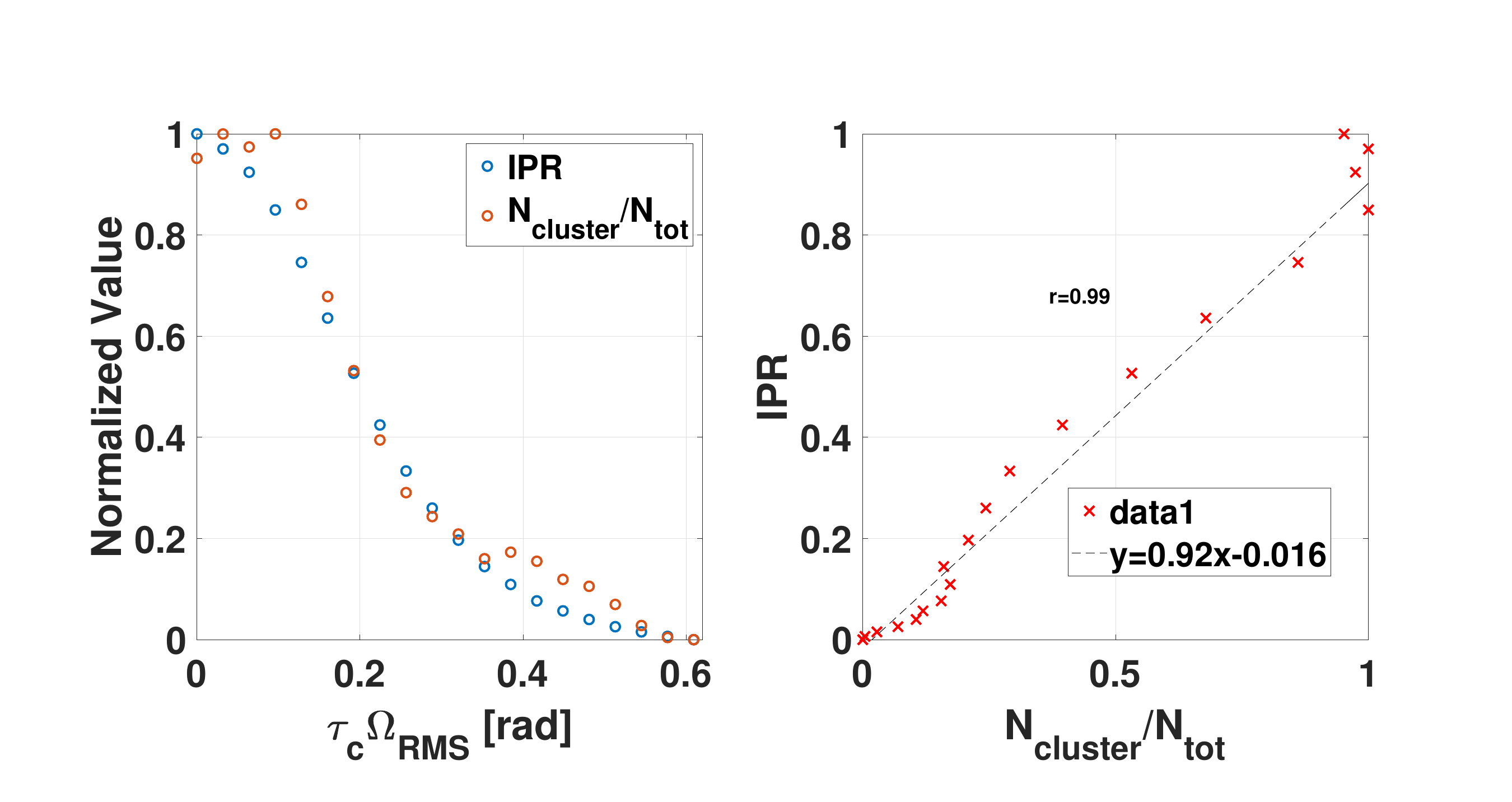}
    \caption{Calculated normalized IPR (blue) and normalized average synchronized cluster size (red) as a function of $\tau_c\Omega_{RMS}$, and IPR vs. synchronized cluster size. The results are highly correlated as indicated by the linear fit and correlation coefficient $r = 0.99$.}
    \label{fig_compare}
\end{figure}

In order to further validate that the IPR can be used for quantifying phase synchronization, we performed numerical simulations (using LRE) of a coupled $20\times20$ lasers array where the frequency detuning randomly varied. The detuning disorder causes the array to split into spatially separated clusters of phase synchronized lasers, which decrease in size as the disorder is increased \cite{strogatz1988collective,sakaguchi1987local}. In the simulation we found the frequencies of the lasers in the array from which we determined the average synchronized cluster size. The average synchronized cluster size grows as the number of neighboring lasers with the same frequency increases. Figure \ref{fig_compare} shows a comparison between the average synchronized cluster size in the array and the FF IPR. Our results show that the results of the two methods are very well correlated as expected, and so it is reasonable to use the FF IPR to quantify the average synchronized cluster size.
\newpage
\section{Correlated disorder results for weak coupling}
\begin{figure}[H]
    \centering
    \includegraphics[trim=0cm  0cm 0cm 0cm,clip,width=\linewidth]{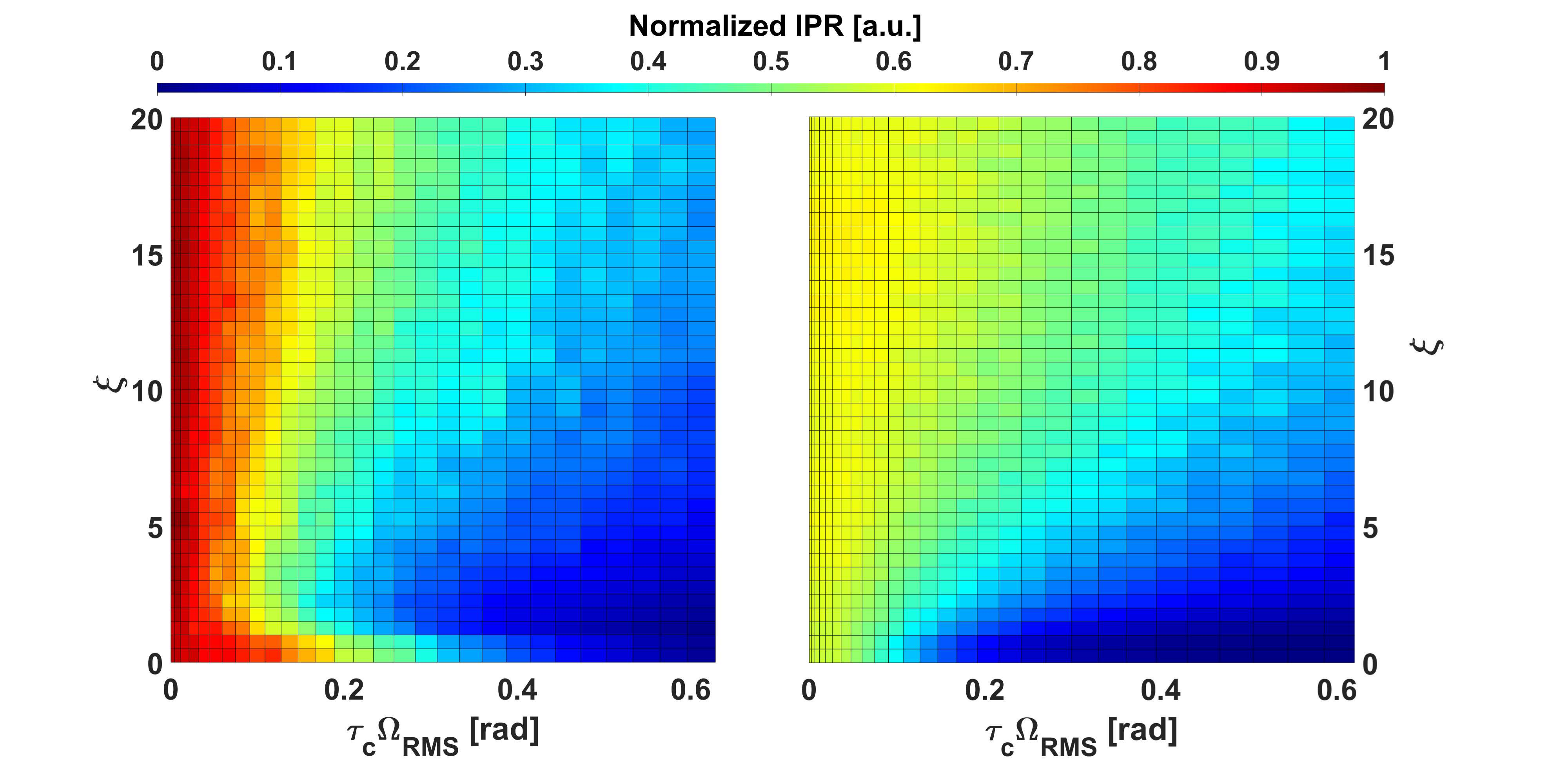}
    \caption{Measured IPR as a function of $\tau_c\Omega_{RMS},\xi$, for $\abs{K}= 0.25 \text{(left)}$, and for $\abs{K} = 0.12 \text{(right)}$}
    \label{fig_phasediagweak}
\end{figure}

\begin{figure}[H]
    \centering
    \includegraphics[trim=2cm  3cm 1cm 2cm,clip,width=0.7\linewidth]{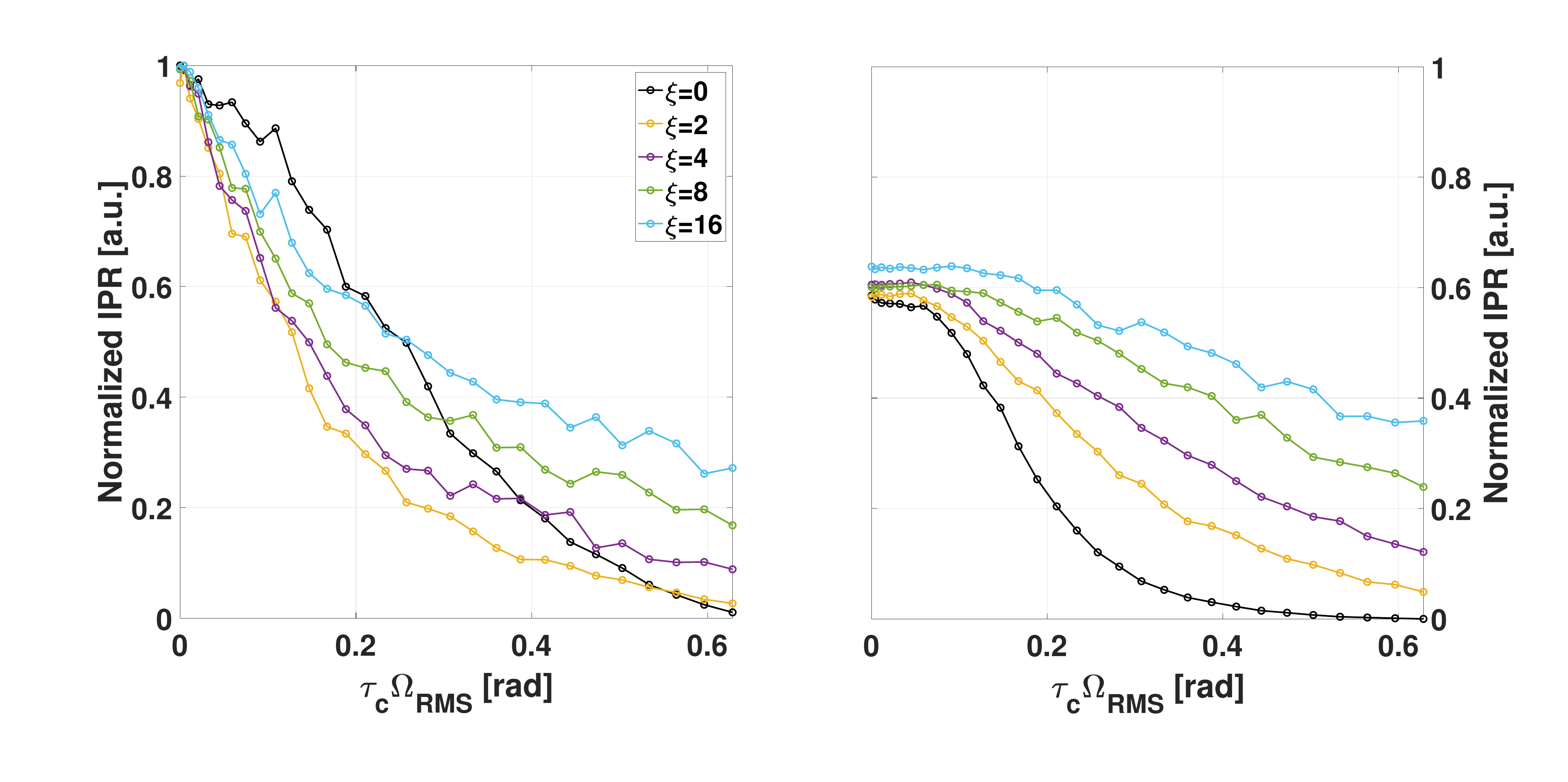}
    \caption{Experimentally measured normalized FF IPR as a function of disorder, $\tau_c\Omega_{RMS}$ for $\abs{K}=0.25$ (left) and $|K|=0.12$ (right). Different colored plots correspond to different applied correlation lengths $\xi$.}
    \label{fig_weak_det}
\end{figure}

\begin{figure}[H]
    \centering
    \includegraphics[trim=3cm  3cm 0cm 2cm,clip,width=0.7\linewidth]{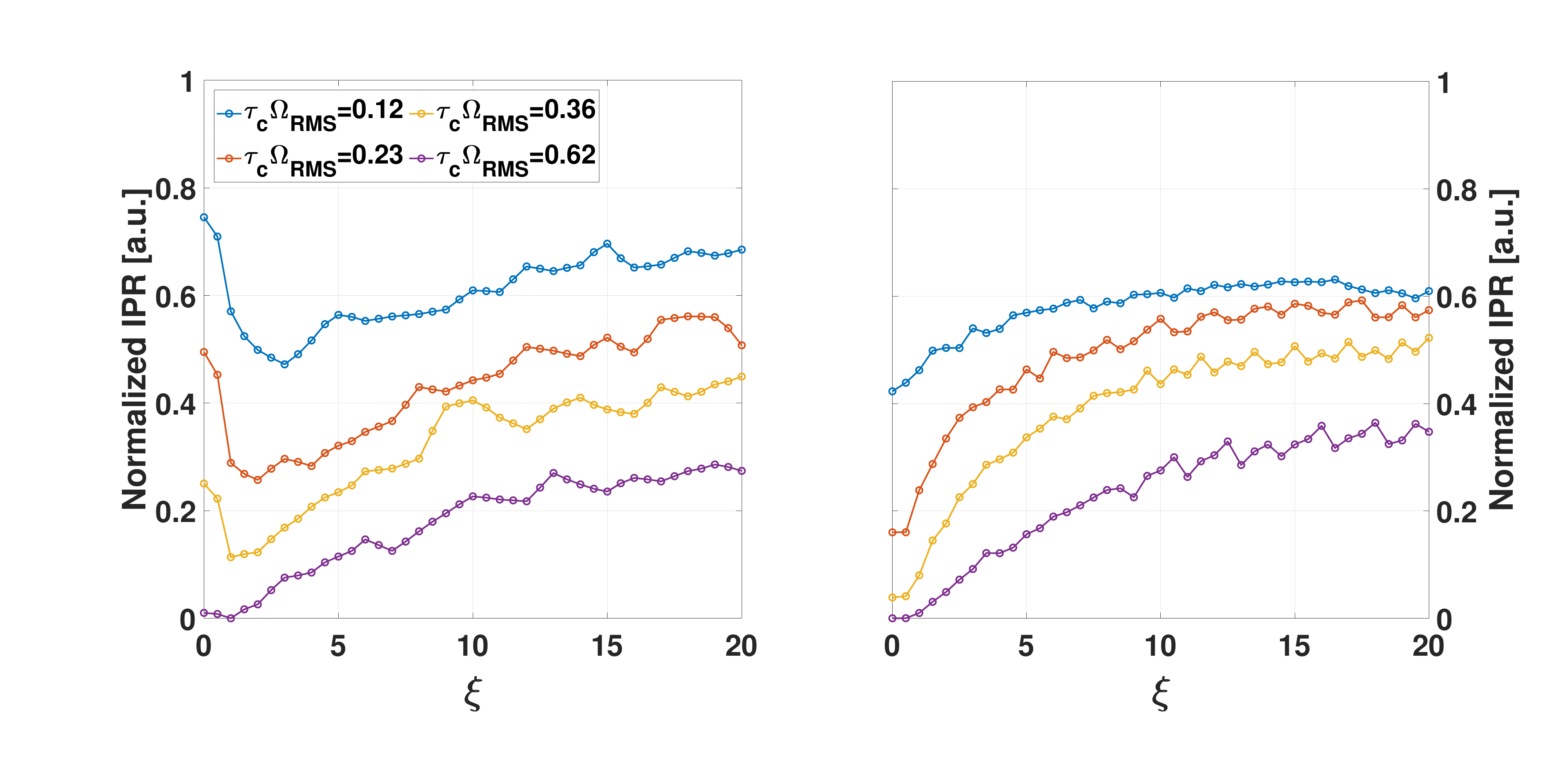}
    \caption{Experimentally measured normalized FF IPR as a function of correlation length $\xi$ for  $\abs{K}=0.25$ (left) and $\abs{K}=0.12$ (right). Different colored plots correspond to different magnitudes of applied disorder, $\tau_c\Omega_{RMS}$.}
    \label{fig_weak_xi}
\end{figure}

We repeated the experiment whose measurements are shown in Fig. 3-5 in the main text with a weaker coupling strength of $\abs{K}=0.12$. The results of both experiments are presented side by side in Figures \ref{fig_phasediagweak}-\ref{fig_weak_xi}. The IPR values presented are normalized such that $IPR=1,0$ are the largest and smallest IPR values measured across all experiments, respectively.  It is interesting to note that the non-monotonic behavior in $\xi$ is absent with the weak coupling, and an increase of the disorder correlation length causes a slower decay of the IPR. This is in agreement with our theoretical and numerical analysis: Since the initial synchronized cluster size is smaller in the case of weak coupling, all measured values of $\xi$ are at the $\xi\le N$ regime. As a result, disorder with any non-zero correlation length will cause a slower decay of the IPR compared to the uncorrelated case.

\newpage
\section{Derivation of $\ev{\abs{X_j}}^2$ in Eq.(5)}\label{strogatz_param_derivation}

To derive the explicit analytical expression presented in Eq.(6)
, we start with a series of $L$ independent normal random variables,
$$Z_i=\mathcal{N}(0,\sigma_Z^2),$$

and then construct a series of correlated random variables by convolving the $Z$ variables with a Gaussian with waist $\xi$, to yield:
$$r_j=\sum_{i=0}^{L}e^{-\frac{(i-j)^2}{\xi^2}}Z_i$$

We note that $X_j$, which is the accumulated deviation from the mean defined in Eq.(4) in the main text, can be expressed in terms of partial sums of the series,
$$X_j(N,\xi)=\sum_{i=0}^j(r_i-\bar{r})=S_j(\xi)-\frac{j}{N}S_N(\xi),$$
where we define the partial sum $S_n$ as:
$$S_n\equiv \sum_{j=0}^n r_j$$

We now write the value of $S_n$ explicitly:
$$S_n=\sum_{j=0}^nr_j=\sum_{j=0}^n\sum_{i=0}^Le^{-\frac{(i-j)^2}{\xi^2}}Z_i,$$
and insert it into $X_n$:
$$X_n=S_n-\frac{n}{N}S_N=\sum_{i=0}^L[\sum_{j=0}^n e^{-\frac{(i-j)^2}{\xi^2}}Z_i-\frac{n}{N}\sum_{k=0}^Ne^{-\frac{(i-k)^2}{\xi^2}}Z_i]$$

we approximate the sums as integrals using the first order Euler-Maclaurin approximation:

$$\approx \sum_{i=0}^L[\int_0^n e^{-\frac{(j-i)^2}{\xi^2}}\dd{j}+\frac{e^{-\frac{(n-i)^2}{\xi^2}}-e^{-\frac{i^2}{\xi^2}}}{2}-\frac{n}{N}\int_0^Ne^{-\frac{(i-k)^2}{\xi^2}}\dd{k}-\frac{n}{N}\frac{e^{-\frac{(N-i)^2}{\xi^2}}-e^{-\frac{i^2}{\xi^2}}}{2}]Z_i$$
$$=\sum_{i=0}^L[\operatorname{erf}(\frac{n-i}{\xi})+\operatorname{erf}(\frac{i}{\xi})+\frac{e^{-\frac{(n-i)^2}{\xi^2}}-e^{-\frac{i^2}{\xi^2}}}{2}-\frac{n}{N}\operatorname{erf}(\frac{N-i}{\xi})-\frac{n}{N}\operatorname{erf}(\frac{i}{\xi})-\frac{n}{N}\frac{e^{-\frac{(N-i)^2}{\xi^2}}-e^{-\frac{i^2}{\xi^2}}}{2}]\frac{\sqrt{\pi}}{2}\xi Z_i$$

For $n\rightarrow N$, we expect $X_n\rightarrow 0$ since the partial and total sum are identical, and indeed we see that it is the case.

At this point we note that as a result of the method we used to generate the correlated variables, $r_n$, we effectively changed their standard deviation. Therefore, we need to calculate the new standard deviation $\sigma_L$ and normalize by it to ensure the $r_n$ variables have the required standard deviation. For a single variable we get:
$$E[r_n^2]=E[(\sum_{i=0}^Le^{-\frac{(i-n)^2}{\xi^2}}Z_i)^2]=\sum_{i=0}^Le^{-\frac{2(i-n)^2}{\xi^2}}E[Z_i^2]=\frac{\sqrt{\pi}}{2\sqrt{2}}\xi\sigma_Z^2[\text{erf}(\frac{\sqrt{2}n}{\xi})+\text{erf}(\frac{\sqrt{2}(L-n)}{\xi}]$$
where we used the fact that the $Z_i$ variables are independent, so $E[Z_iZ_j]=\delta_{ij}$.
For the entire series of $L$ variables we get:
$$\sigma_L^2=\frac{\sqrt{\pi}}{2\sqrt{2}}\frac{\xi\sigma_Z^2}{L}\sum_{n=0}^L[\text{erf}(\frac{\sqrt{2}n}{\xi})+\text{erf}(\frac{\sqrt{2}(L-n)}{\xi}]\approx\frac{2\xi\sigma_Z^2}{L}\int_0^L\text{erf}(\frac{\sqrt{2}{n}}{\xi})\dd{n}$$
Using the known integral solution,
$$\int_0^x a\cdot\text{erf}(\frac{(x'-b)}{a})\dd{x}=\frac{a^2}{\sqrt{\pi}}(e^{-\frac{(x-b)^2}{a^2}}-e^{-\frac{b^2}{a^2}})+a(x-b)\cdot\text{erf}(\frac{x-b}{a})-ab\cdot\text{erf}(\frac{b}{a})$$

yields
$$\sigma_L^2 = \frac{2\xi\sigma_Z^2}{L}[\frac{\xi}{\sqrt{2\pi}}(e^{-\frac{2L^2}{\xi^2}}-1)+L\cdot\text{erf}(\frac{\sqrt{2}L}{\xi})]$$

Finally, we use the usual properties of normal variables and normalize by $\sigma_L^2$ to get that:
$$\ev{\abs{X_n}}^2\approx \frac{\pi \xi L}{8}\frac{\sum_{i=0}^L[\operatorname{erf}(\frac{n-i}{\xi})+\frac{1}{2}e^{-\frac{(n-i)^2}{\xi^2}}+(1-\frac{n}{N})(\operatorname{erf}(\frac{i}{\xi})-\frac{1}{2}e^{-\frac{i^2}{\xi^2}})-\frac{n}{N}\cdot\operatorname{erf}(\frac{N-i}{\xi})]^2-\frac{1}{2}e^{-\frac{(N-i)^2}{\xi^2}})}{\frac{\xi}{\sqrt{2\pi}}(e^{-\frac{2L^2}{\xi^2}}-1)+L\cdot\operatorname{erf}(\frac{\sqrt{2}L}{\xi})}$$

\bibliographystyle{unsrt}
\bibliography{main}

\begin{thebibliography}{10}

\bibitem{kiely2018relationship}
Thomas~G Kiely and J.~K. Freericks.
\newblock Relationship between the transverse-field ising model and the x y model via the rotating-wave approximation.
\newblock {\em Physical Review A}, 97(2):023611, 2018.

\bibitem{henkel1984statistical}
Malte Henkel.
\newblock Statistical mechanics of the 2d quantum xy model in a transverse field.
\newblock {\em Journal of Physics A: Mathematical and General}, 17(14):L795, 1984.

\bibitem{juhasz2014random}
R{\'o}bert Juh{\'a}sz, Istv{\'a}n~A Kov{\'a}cs, and Ferenc Igl{\'o}i.
\newblock Random transverse-field ising chain with long-range interactions.
\newblock {\em Europhysics Letters}, 107(4):47008, 2014.

\bibitem{mckenzie1996exact}
Ross~H McKenzie.
\newblock Exact results for quantum phase transitions in random xy spin chains.
\newblock {\em Physical Review Letters}, 77(23):4804, 1996.

\bibitem{fisher1992random}
Daniel~S Fisher.
\newblock Random transverse field ising spin chains.
\newblock {\em Physical Review Letters}, 69(3):534, 1992.

\bibitem{acebron2005kuramoto}
Juan~A Acebr{\'o}n, Luis~L Bonilla, Conrad J~P{\'e}rez Vicente, F{\'e}lix Ritort, and Renato Spigler.
\newblock The kuramoto model: A simple paradigm for synchronization phenomena.
\newblock {\em Reviews of Modern Physics}, 77(1):137, 2005.

\bibitem{wiesenfeld1998frequency}
Kurt Wiesenfeld, Pere Colet, and Steven~H Strogatz.
\newblock Frequency locking in josephson arrays: Connection with the kuramoto model.
\newblock {\em Physical Review E}, 57(2):1563, 1998.

\bibitem{trees2005synchronization}
Brad~R Trees, Vinod Saranathan, and David Stroud.
\newblock Synchronization in disordered josephson junction arrays: Small-world connections and the kuramoto model.
\newblock {\em Physical Review E}, 71(1):016215, 2005.

\bibitem{yeung1999time}
M.~K.~Stephen Yeung and Steven~H Strogatz.
\newblock Time delay in the kuramoto model of coupled oscillators.
\newblock {\em Physical Review Letters}, 82(3):648, 1999.

\bibitem{takemura2021emulating}
Naotomo Takemura, Kenta Takata, Masato Takiguchi, and Masaya Notomi.
\newblock Emulating the local kuramoto model with an injection-locked photonic crystal laser array.
\newblock {\em Scientific Reports}, 11(1):8587, 2021.

\bibitem{shahal2020synchronization}
Shir Shahal, Ateret Wurzberg, Inbar Sibony, Hamootal Duadi, Elad Shniderman, Daniel Weymouth, Nir Davidson, and Moti Fridman.
\newblock Synchronization of complex human networks.
\newblock {\em Nature Communications}, 11(1):3854, 2020.

\bibitem{niederberger2010disorder}
Armand Niederberger, Marek~M Rams, Jacek Dziarmaga, Fernando~M Cucchietti, Jan Wehr, and Maciej Lewenstein.
\newblock Disorder-induced order in quantum xy chains.
\newblock {\em Physical Review A}, 82(1):013630, 2010.

\bibitem{villain1980order}
Jacques Villain, R~Bidaux, J-P Carton, and R~Conte.
\newblock Order as an effect of disorder.
\newblock {\em Journal de Physique}, 41(11):1263--1272, 1980.

\bibitem{sonnenschein2013approximate}
Bernard Sonnenschein and Lutz Schimansky-Geier.
\newblock Approximate solution to the stochastic kuramoto model.
\newblock {\em Physical Review E}, 88(5):052111, 2013.

\bibitem{granato1986quenched}
Enzo Granato and J.~M. Kosterlitz.
\newblock Quenched disorder in josephson-junction arrays in a transverse magnetic field.
\newblock {\em Physical Review B}, 33(9):6533, 1986.

\bibitem{vlasov2013synchronization}
Vladimir Vlasov and Arkady Pikovsky.
\newblock Synchronization of a josephson junction array in terms of global variables.
\newblock {\em Physical Review E}, 88(2):022908, 2013.

\bibitem{rouzaire2021defect}
Ylann Rouzaire and Demian Levis.
\newblock Defect superdiffusion and unbinding in a 2d xy model of self-driven rotors.
\newblock {\em Physical Review Letters}, 127(8):088004, 2021.

\bibitem{contractor2022scalable}
Rushin Contractor, Wanwoo Noh, Walid Redjem, Wayesh Qarony, Emma Martin, Scott Dhuey, Adam Schwartzberg, and Boubacar Kant{\'e}.
\newblock Scalable single-mode surface-emitting laser via open-dirac singularities.
\newblock {\em Nature}, 608(7924):692--698, 2022.

\bibitem{alex2021}
Alex Dikopoltsev, Tristan~H. Harder, Eran Lustig, Oleg~A. Egorov, Johannes Beierlein, Adriana Wolf, Yaakov Lumer, Monika Emmerling, Christian Schneider, Sven Höfling, Mordechai Segev, and Sebastian Klembt.
\newblock Topological insulator vertical-cavity laser array.
\newblock {\em Science}, 373(6562):1514--1517, 2021.

\bibitem{bandres2018topological}
Miguel~A Bandres, Steffen Wittek, Gal Harari, Midya Parto, Jinhan Ren, Mordechai Segev, Demetrios~N Christodoulides, and Mercedeh Khajavikhan.
\newblock Topological insulator laser: Experiments.
\newblock {\em Science}, 359(6381):eaar4005, 2018.

\bibitem{rosiek2023observation}
Christian~Anker Rosiek, Guillermo Arregui, Anastasiia Vladimirova, Marcus Albrechtsen, Babak Vosoughi~Lahijani, Rasmus~Elleb{\ae}k Christiansen, and S{\o}ren Stobbe.
\newblock Observation of strong backscattering in valley-hall photonic topological interface modes.
\newblock {\em Nature Photonics}, 17(5):386--392, 2023.

\bibitem{wang2009observation}
Zheng Wang, Yidong Chong, John~D Joannopoulos, and Marin Solja{\v{c}}i{\'c}.
\newblock Observation of unidirectional backscattering-immune topological electromagnetic states.
\newblock {\em Nature}, 461(7265):772--775, 2009.

\bibitem{hafezi2011robust}
Mohammad Hafezi, Eugene~A Demler, Mikhail~D Lukin, and Jacob~M Taylor.
\newblock Robust optical delay lines with topological protection.
\newblock {\em Nature Physics}, 7(11):907--912, 2011.

\bibitem{vishwatopo}
Vishwa Pal, Chene Tradonsky, Ronen Chriki, Asher~A. Friesem, and Nir Davidson.
\newblock Observing dissipative topological defects with coupled lasers.
\newblock {\em Physical Review Letters}, 119:013902, Jul 2017.

\bibitem{yang2022topological}
Lechen Yang, Guangrui Li, Xiaomei Gao, and Ling Lu.
\newblock Topological-cavity surface-emitting laser.
\newblock {\em Nature Photonics}, 16(4):279--283, 2022.

\bibitem{ruter2010observation}
Christian~E R{\"u}ter, Konstantinos~G Makris, Ramy El-Ganainy, Demetrios~N Christodoulides, Mordechai Segev, and Detlef Kip.
\newblock Observation of parity--time symmetry in optics.
\newblock {\em Nature Physics}, 6(3):192--195, 2010.

\bibitem{arwas2022anyonic}
Geva Arwas, Sagie Gadasi, Igor Gershenzon, Asher Friesem, Nir Davidson, and Oren Raz.
\newblock Anyonic-parity-time symmetry in complex-coupled lasers.
\newblock {\em Science Advances}, 8(22), 2022.

\bibitem{nixon2013}
Micha Nixon, Eitan Ronen, Asher~A Friesem, and Nir Davidson.
\newblock Observing geometric frustration with thousands of coupled lasers.
\newblock {\em Physical Review Letters}, 110(18):184102, 2013.

\bibitem{wang2013coherent}
Zhe Wang, Alireza Marandi, Kai Wen, Robert~L Byer, and Yoshihisa Yamamoto.
\newblock Coherent ising machine based on degenerate optical parametric oscillators.
\newblock {\em Physical Review A}, 88(6):063853, 2013.

\bibitem{mcmahon2016fully}
Peter~L McMahon, Alireza Marandi, Yoshitaka Haribara, Ryan Hamerly, Carsten Langrock, Shuhei Tamate, Takahiro Inagaki, Hiroki Takesue, Shoko Utsunomiya, Kazuyuki Aihara, et~al.
\newblock A fully programmable 100-spin coherent ising machine with all-to-all connections.
\newblock {\em Science}, 354(6312):614--617, 2016.

\bibitem{marandi2014network}
Alireza Marandi, Zhe Wang, Kenta Takata, Robert~L Byer, and Yoshihisa Yamamoto.
\newblock Network of time-multiplexed optical parametric oscillators as a coherent ising machine.
\newblock {\em Nature Photonics}, 8(12):937--942, 2014.

\bibitem{tradonsky2019rapid}
C~Tradonsky, I~Gershenzon, V~Pal, R~Chriki, AA~Friesem, O~Raz, and N~Davidson.
\newblock Rapid laser solver for the phase retrieval problem.
\newblock {\em Science Advances}, 5(10), 2019.

\bibitem{Pando:23}
Amit Pando, Sagie Gadasi, Asher Friesem, and Nir Davidson.
\newblock Improved laser phase locking with intra-cavity adaptive optics.
\newblock {\em Optics Express}, 31(4):6947--6955, 2023.

\bibitem{Tradonsky:21}
Chene Tradonsky, Simon Mahler, Gaodi Cai, Vishwa Pal, Ronen Chriki, Asher~A. Friesem, and Nir Davidson.
\newblock High-resolution digital spatial control of a highly multimode laser.
\newblock {\em Optica}, 8(6):880--884, Jun 2021.

\bibitem{cao2019complex}
Hui Cao, Ronen Chriki, Stefan Bittner, Asher~A Friesem, and Nir Davidson.
\newblock Complex lasers with controllable coherence.
\newblock {\em Nature Reviews Physics}, 1(2):156--168, 2019.

\bibitem{Arnaud:69}
JA~Arnaud.
\newblock Degenerate optical cavities.
\newblock {\em Applied Optics}, 8(1):189--196, 1969.

\bibitem{tradonsky2017talbot}
Chene Tradonsky, Vishwa Pal, Ronen Chriki, Nir Davidson, and Asher~A Friesem.
\newblock Talbot diffraction and fourier filtering for phase locking an array of lasers.
\newblock {\em Applied Optics}, 56(1):A126--A132, 2017.

\bibitem{talbotLeger}
James~R. Leger.
\newblock Lateral mode control of an algaas laser array in a talbot cavity.
\newblock {\em Applied Physics Letters}, 55(4):334--336, 1989.

\bibitem{cohen2009single}
Sharona~Sedghani Cohen, Vardit Eckhouse, Asher~A Friesem, and Nir Davidson.
\newblock Single frequency lasing using coherent combining.
\newblock {\em Optics Communications}, 282(9):1861--1866, 2009.

\bibitem{fridman2010phase}
Moti Fridman, Micha Nixon, Eitan Ronen, Asher~A Friesem, and Nir Davidson.
\newblock Phase locking of two coupled lasers with many longitudinal modes.
\newblock {\em Optics Letters}, 35(4):526--528, 2010.

\bibitem{smith1965stabilized}
P~Smith.
\newblock Stabilized, single-frequency output from a long laser cavity.
\newblock {\em IEEE Journal of Quantum Electronics}, 1(8):343--348, 1965.

\bibitem{Supplemental}
See Supplemental Material at [Link] for detailed experimental arrangement, equations derivations, and additional experimental results.

\bibitem{mandel1965coherence}
Leonard Mandel and Emil Wolf.
\newblock Coherence properties of optical fields.
\newblock {\em Reviews of Modern Physics}, 37(2):231, 1965.

\bibitem{friberg1983spatial}
Ari~T Friberg and Ronald~J Sudol.
\newblock The spatial coherence properties of gaussian schell-model beams.
\newblock {\em Optica Acta: International Journal of Optics}, 30(8):1075--1097, 1983.

\bibitem{Marc_ipr}
Dirk Witthaut and Marc Timme.
\newblock Kuramoto dynamics in hamiltonian systems.
\newblock {\em Physical Review E}, 90(3):032917, 2014.

\bibitem{longhi2022non}
Stefano Longhi.
\newblock Non-hermitian laser arrays with tunable phase locking.
\newblock {\em Optics Letters}, 47(8):2040--2043, 2022.

\bibitem{pal2020rapid}
Vishwa Pal, Simon Mahler, Chene Tradonsky, Asher~A Friesem, and Nir Davidson.
\newblock Rapid fair sampling of the x y spin hamiltonian with a laser simulator.
\newblock {\em Physical Review Research}, 2(3):033008, 2020.

\bibitem{chriki2018spatiotemporal}
Ronen Chriki, Simon Mahler, Chene Tradonsky, Vishwa Pal, Asher~A Friesem, and Nir Davidson.
\newblock Spatiotemporal supermodes: Rapid reduction of spatial coherence in highly multimode lasers.
\newblock {\em Physical Review A}, 98(2):023812, 2018.

\bibitem{PhysRevLett.92.093905}
Fabien Rogister, K~Scott Thornburg~Jr, Larry Fabiny, Michael M{\"o}ller, and Rajarshi Roy.
\newblock Power-law spatial correlations in arrays of locally coupled lasers.
\newblock {\em Physical Review Letters}, 92(9):093905, 2004.

\bibitem{honari2020mapping}
Mostafa Honari-Latifpour and Mohammad-Ali Miri.
\newblock Mapping the x y hamiltonian onto a network of coupled lasers.
\newblock {\em Physical Review Research}, 2(4):043335, 2020.

\bibitem{strogatz1988collective}
Steven~H Strogatz and Renato~E Mirollo.
\newblock Collective synchronisation in lattices of nonlinear oscillators with randomness.
\newblock {\em Journal of Physics A: Mathematical and General}, 21(13):L699, 1988.

\bibitem{sakaguchi1987local}
Hidetsugu Sakaguchi, Shigeru Shinomoto, and Yoshiki Kuramoto.
\newblock Local and grobal self-entrainments in oscillator lattices.
\newblock {\em Progress of Theoretical Physics}, 77(5):1005--1010, 1987.

\end{thebibliography}

\end{document}